\documentclass[journal]{IEEEtran}
\usepackage{cite}

\ifCLASSINFOpdf
\else
\fi
\usepackage{array}

\usepackage{comment}
\usepackage[hidelinks]{hyperref}
\usepackage{booktabs}
\usepackage[dvipsnames,table]{xcolor}
\usepackage{multirow}

\usepackage{graphicx}
\usepackage{tikz}
\usetikzlibrary{arrows}
\usetikzlibrary{fit}
\usetikzlibrary{patterns}
\usetikzlibrary{calc}
\usepackage{xpatch}
\usepackage{pgfplots}
\usepackage{paralist}
\usepackage{amssymb}
\usepackage{wrapfig}

\usepackage{pifont}
\newcommand{\cmark}{\ding{51}}%
\newcommand{\xmark}{{\ding{55}}}%
\newcommand{\gcmark}{\textcolor{green}{\cmark}}
\newcommand{\rxmark}{\textcolor{red}{\xmark}}

\newcommand{\eg}{{e.g.}}
\newcommand{\ie}{{\textit{i.e.}}}
\newcommand{\etal}{{et al.}}

\newcommand{\attackSection}[1]{{\setlength{\parskip}{.5\baselineskip plus2pt minus2pt}\textbf{#1.}}}
\newcommand{\attacks}{{\underline{Attacks}. }}

\newif\ifrefsinfig
 \refsinfigtrue
 
\newif\ifrefsintab
 \refsintabtrue
 
\newif\ifpreprint
 \preprintfalse

\makeatletter
\def\ps@IEEEtitlepagestyle{
  \def\@oddfoot{\mycopyrightnotice}
  \def\@evenfoot{}
}
\def\mycopyrightnotice{
  {\footnotesize
  \hspace*{-.2cm}\fbox{\begin{minipage}{\textwidth}
  \centering
  Copyright~\copyright~2017 IEEE. Personal use of this material is permitted. Permission from IEEE must be obtained for all other uses, in any current or future media, including reprinting/republishing this material for advertising or promotional purposes, creating new collective works, for resale or redistribution to servers or lists, or reuse of any copyrighted component of this work in other works. The original IEEE publication is available at: \url{https://doi.org/10.1109/COMST.2017.2779824}
  \end{minipage}}
  }
}

\PassOptionsToPackage{hidelinks}{hyperref}
\begin{document}
%
\title{Systematic Classification of Side-Channel Attacks: A Case Study for Mobile Devices}
%
%
%

\author{Raphael~Spreitzer, Veelasha Moonsamy, Thomas Korak, and Stefan Mangard%
\thanks{R. Spreitzer, and S. Mangard are with
Graz University of Technology, IAIK, Graz, Austria. (e-mail: raphael.spreitzer@iaik.tugraz.at, stefan.mangard@iaik.tugraz.at).}
\thanks{T. Korak was with Graz University of Technology, IAIK, Graz, Austria.}
\thanks{V. Moonsamy is with Radboud University, Digital Security Group,
Nijmegen, The Netherlands. (e-mail: email@veelasha.org).}
\ifpreprint\thanks{\copyright 2017 IEEE. Personal use of this material is permitted. Permission from IEEE must be obtained for all other uses, in any current or future media, including reprinting/republishing this material for advertising or promotional purposes, creating new collective works, for resale or redistribution to servers or lists, or reuse of any copyrighted component of this work in other works.}\fi
}

%
%

\markboth{IEEE Communications Surveys \& Tutorials,~Vol.~XX, No.~Z, Month~YYYY}%
{Spreitzer \MakeLowercase{\textit{et al.}}: Systematic Classification of Side-Channel Attacks: A Case Study for Mobile Devices}

%



\maketitle

\begin{abstract}
Side-channel attacks on mobile devices have gained increasing attention since their introduction in 2007. While traditional side-channel attacks, such as power analysis attacks and electromagnetic analysis attacks, required physical presence of the attacker as well as expensive equipment, an (unprivileged) application is all it takes to exploit the leaking information on modern mobile devices. Given the vast amount of sensitive information that are stored on smartphones, the ramifications of side-channel attacks affect both the security and privacy of users and their devices. 

In this paper, we propose a new categorization system for side-channel attacks, which is necessary as side-channel attacks have evolved significantly since their scientific investigations during the smart card era in the 1990s. Our proposed classification system allows to analyze side-channel attacks systematically, and facilitates the development of novel countermeasures. Besides this new categorization system, the extensive survey of existing attacks and attack strategies provides valuable insights into the evolving field of side-channel attacks, especially when focusing on mobile devices. We conclude by discussing open issues and challenges in this context and outline possible future research directions.
\end{abstract}


%
\IEEEpeerreviewmaketitle

\section{Introduction}
 \label{sec:intro}
%
%
%
%
\IEEEPARstart{S}{ide}-channel attacks exploit (unintended) information leakage of computing devices or implementations to infer sensitive information. Starting with the seminal works of Kocher~\cite{DBLP:conf/crypto/Kocher96}, Kocher~\etal~\cite{DBLP:conf/crypto/KocherJJ99}, Quisquater and Samyde~\cite{DBLP:conf/esmart/QuisquaterS01}, as well as Mangard~\etal~\cite{DBLP:books/daglib/0017272}, many follow-up papers considered attacks against cryptographic implementations to exfiltrate key material from smart cards by means of timing information, power consumption, or electromagnetic (EM) emanation. 
These ``traditional'' side-channel attacks required the attacker to be in physical possession of the device to be able to observe and learn the leaking information, yet different attacks assumed different types of attackers and different levels of invasiveness. More specifically, in order to systematically analyze side-channel attacks, they have been categorized along the following two orthogonal axes:
\begin{enumerate}
 \item \emph{Active} vs \emph{passive}: Depending on whether the attacker actively influences the behavior of the device or only passively observes leaking information. 
 \item \emph{Invasive} vs \emph{semi-invasive} vs \emph{non-invasive}: Depending on whether or not the attacker removes the passivation layer of the chip, depackages the chip, or does not manipulate the packaging at all.  
\end{enumerate}

However, with the era of cloud computing, the scope and the scale of side-channel attacks have changed significantly in the early 2000s. While early attacks required attackers to be in physical possession of the device, newer side-channel attacks such as cache-timing attacks~\cite{DBLP:journals/joc/TromerOS10,DBLP:conf/uss/YaromF14,Ge2016} or DRAM row buffer attacks~\cite{DBLP:conf/uss/PesslGMSM16} are conducted remotely by executing malicious software in the targeted cloud environment. 
With the advent of mobile devices, and in particular the plethora of embedded features and sensors, even more sophisticated side-channel attacks targeting smartphones have been proposed since around the year 2010. For example, attacks allow to infer keyboard input on touchscreens via sensor readings from native apps~\cite{DBLP:conf/uss/CaiC11,DBLP:conf/acsac/AvivSBS12,DBLP:journals/popets/SimonXA16} and websites~\cite{DBLP:journals/istr/MehrnezhadTSH16}, to deduce a user's location via the power consumption available from the proc filesystem (procfs)~\cite{DBLP:conf/uss/MichalevskySVBN15}, and also to infer a user's identity, location, and diseases~\cite{DBLP:conf/ccs/ZhouDHNPWGN13} via the procfs. 

Clearly, side-channel attacks have a long history and have evolved significantly from attacks on specialized computing devices in the smart card era, to attacks on general-purpose computing platforms in desktop computers and cloud computing infrastructures, and finally to attacks on mobile devices. 
Although side-channel attacks and platform security are already well-studied topics, it must be noted that smartphone security and associated privacy aspects differ from platform security in the context of smart cards, desktop computers, and cloud computing. Especially the following \emph{key enablers} enable more devastating attacks on mobile devices. 
\begin{enumerate}
 \item \textit{Always-on and portability}: First and foremost, mobile devices are always turned on and due to their mobility they are carried around at all times. Thus, they are tightly integrated into our everyday lives. 
 \item \textit{Bring your own device (BYOD)}:
 To decrease the number of devices carried around, employees use personal devices to process corporate data and to access corporate infrastructure, which clearly indicates the importance of secure mobile devices. 
 \item \textit{Ease of software installation}: Due to the appification~\cite{DBLP:conf/sp/Acar0BFM016} of mobile devices, \ie, where there is an app for almost everything, additional software can be installed easily by means of established app markets. Hence, malicious apps can also be spread at a fast pace.  
 \item \textit{OS based on Linux kernel}: Modern mobile operating systems (OS), for example, Android, are based on the Linux kernel. The Linux kernel, however, has initially been designed for desktop machines and information or features that are considered harmless on these platforms turn out to be an immense security and/or privacy threat on mobile devices (cf.~\cite{DBLP:conf/sp/ZhangY0ZW15}). 
 \item \textit{Features and sensors}: 
 Last but not least, these devices include many features and sensors, which are not present on traditional platforms. Due to the inherent nature of mobile devices (always-on and carried around, connectivity, inherent input methods, etc.), such features often enable devastating side-channel attacks. 
 Besides, these sensors have also been used to attack external hardware, such as keyboards and computer hard drives~\cite{DBLP:conf/ccs/MarquardtVCT11,DBLP:conf/ccs/ZhuMZL14,DBLP:conf/fc/Biedermann0S15}, to infer videos played on TVs~\cite{DBLP:conf/percom/SchwittmannMWW16}, and even to attack 3D printers~\cite{DBLP:conf/ccs/SongLBRZX16,DBLP:conf/ccs/HojjatiASCNMWGK16}, which clearly demonstrates the immense power of mobile devices.
\end{enumerate}

Due to the above mentioned \emph{key enablers}, a new area of side-channel attacks has evolved and the majority of more recent side-channel attacks are strictly non-invasive and rely on the execution of malicious software in the targeted environment.
Considering these developments, we observe that the classification system that has been established to analyze side-channel attacks on smart cards does not meet these new attack settings and strategies anymore. 
Hence, the existing classification system does not allow a systematic categorization of modern side-channel attacks, including side-channel attacks on mobile devices. 

In this work, we close this gap by establishing a new categorization system for modern side-channel attacks on mobile devices. 
Therefore, we survey existing side-channel attacks and identify commonalities between them. 
The gained insights allow researchers to identify future research directions and to cope with these attacks on a larger scale.

\subsection{Motivation and High-Level Categorization} 
It is important to note that side-channel attacks against smartphones can be launched by attackers who are in physical possession of the devices and also by remote attackers who managed to spread a seemingly innocuous application via an existing app store. In some cases such side-channel attacks can even be launched via websites and, thus, without relying on the user to install an app. Nevertheless, in today's appified software platforms where apps are distributed easily via available app markets, an attack scenario requiring the user to install a seemingly harmless game is entirely practical. 

Interestingly, side-channel attacks on smartphones exploit physical properties as well as software properties. 
A malicious application can exploit the accelerometer sensor~\cite{DBLP:conf/uss/CaiC11,DBLP:conf/acsac/AvivSBS12} (a physical property) in order to attack the user input, which is due to the inherent input method relying on touchscreens. In addition, attacks can also be conducted by exploiting software features (a logical property) provided by the Android API or the mobile OS itself (cf.~\cite{DBLP:conf/uss/MichalevskySVBN15,DBLP:conf/ccs/ZhouDHNPWGN13}). This clearly indicates that smartphones significantly broaden the scope as well as the scale of attacks. 
Especially the appification~\cite{DBLP:conf/sp/Acar0BFM016} of mobile platforms---\ie, where there is an app for everything---allows to easily target devices and users at an unprecedented scale compared to the smart card and the cloud setting. 

\begin{figure}
 \centering
  \resizebox{.9\columnwidth}{!}{
 \begin{tikzpicture}[font=\Large]
  \begin{scope}[every node/.style={%
    text width=3.5cm,text depth=3cm,inner sep = 2mm,
    minimum height=3cm,minimum width=3.5cm,
    align=left}]

    \node[](T1){};
    \node[anchor=north west](T2) at (T1.north east){};
    \node[anchor=north west](T3) at (T1.south west){};
    \node[anchor=north west](T4) at (T1.south east){}; 
  \end{scope}

 \begin{scope}[align=center]
  \node [text width=6cm,above=3ex,anchor=south] (T5) at (T1.north east)
   {\textbf{\textsc{What?}\\Hardware (physical)}};

  \node [text width=6cm,below=3ex,anchor=north] (T6) at (T3.south east) {%
    \textbf{Software (logical)}};

  \node [text width=3cm,left=3ex,anchor=east](T7) at (T1.south west) 
   {\textbf{\textsc{How?}\\Physical presence\\(local)}};

  \node [text width=3cm,right=3ex,anchor=west](T8) at (T2.south east) {%
   \textbf{Software\\only\\(remote)}};
 \end{scope}

  \draw[thick,dashed,gray!60] (T5) -- (T6)
                               (T7) -- (T8) ;   
             
  \def\firstellipse{(0,-.2) ellipse [x radius=1.2, y radius=2.1, rotate=-60]}
  \def\secondellipse{(3.2,-1.5) ellipse [x radius=2.9, y radius=1.2, rotate=0]}
  \def\thirdellipse{(2.5,-1) ellipse [x radius=2.2, y radius=4.1, rotate=50]}
  \fill[white,postaction={pattern=crosshatch, pattern color=red!60!white}] \firstellipse; 
  \fill[white,postaction={pattern=north east lines, pattern color=green!100!white}] \secondellipse;
  \fill[NavyBlue!85!white,fill opacity=0.4] \thirdellipse;
  \fill[white,postaction={pattern=crosshatch, pattern color=red!60!white}]   (-3.4,-4) rectangle (-2.9,-3.5) node[pos=0.5,right,black] (R1) {~~~Smart cards};
  \fill[white,postaction={pattern=north east lines, pattern color=green!100!white}] (-3.4,-4.6) rectangle (-2.9,-4.1) node[pos=0.5,right,black]{~~~Cloud};
  \fill[NavyBlue!85!white,fill opacity=.4] (-3.4,-5.2) rectangle (-2.9,-4.7) node[pos=0.5,right,black, fill opacity=1] (R3) {~~~Smartphones};
  \draw (-3.6,-3.3) -| (.5,-5.4) -| (-3.6,-3.3);
\end{tikzpicture} 
}
\caption{Scope of attacks for smart cards, cloud infrastructures, and smartphones.}
\label{fig:dimensions}
\end{figure}
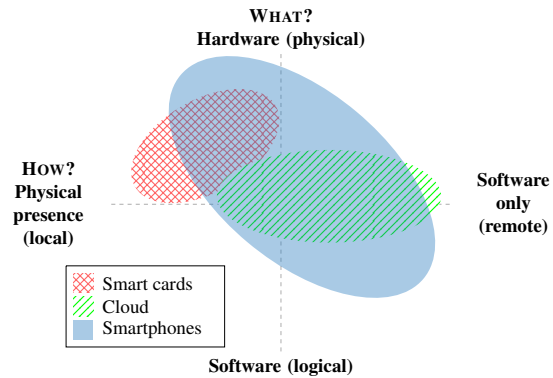

Figure~\ref{fig:dimensions} illustrates a high-level categorization system for side-channel attacks. 
We indicate the type of information that is exploited (\textsc{What?}) and how the adversary learns the leaking information (\textsc{How?}) on the y-axis and x-axis, respectively. 
Furthermore, we indicate how \emph{existing} side-channel attacks against smart cards, cloud computing infrastructures, and smartphones relate to it, \ie, where existing attacks on the respective platforms are located in this new categorization system.
For example, attackers exploit hardware-based information leakage (physical properties) \cite{DBLP:books/daglib/0017272} of smart cards by measuring the power consumption with an oscilloscope. 
In this case, the attacker must be in possession of the device under attack, which is indicated by the red cross-hatched area. 

In contrast, side-channel attacks against cloud-computing infrastructures do not (necessarily) require the attacker to be physically present---unless we consider a malicious cloud provider---as the attacker is able to remotely execute software. Usually, these attacks exploit microarchitectural behavior (such as cache attacks~\cite{DBLP:journals/joc/TromerOS10,DBLP:conf/uss/YaromF14,DBLP:conf/uss/GrussSM15,Ge2016}) or software features (such as page deduplication~\cite{DBLP:conf/esorics/GrussBM15}) in order to infer secret information from co-located processes. Hence, the green dashed area in Figure~\ref{fig:dimensions} is shifted to the right as these attacks mostly rely on software execution, and it is also shifted to the area below the x-axis as these attacks also target software features. 

Even more manifold and diverse side-channel attacks have been proposed for smartphones, which is indicated by the larger area in Figure~\ref{fig:dimensions}. These manifold side-channel attacks mainly result from the five aforementioned \emph{key enablers}. 
More specifically, this area indicates that on smartphones we have to deal with local attackers that exploit physical properties, but also with attackers that execute software on the smartphone in order to exploit both physical properties as well as software features (logical properties, such as the memory footprint~\cite{DBLP:conf/sp/JanaS12a} or the data-usage statistics~\cite{DBLP:conf/ccs/ZhouDHNPWGN13,DBLP:conf/wisec/SpreitzerGKM16}). 
In the remainder of this paper we will refine this high-level categorization system in order to systematically analyze modern side-channel attacks.

Although we do not explicitly focus on Android in this paper, the majority of the existing papers deal with the Android operating system. 
This reflects the trend that the research community focuses mostly on Android because of its openness and also because it has the biggest market share among all mobile operating systems. 
Gartner~\cite{Gartner2017} reports that Android sales (86\% in Q1 2017) clearly outperform Apple iOS sales (14\% in Q1 2017). 

\subsection{Outline}
The remainder of this paper is organized as follows. 
Section~\ref{sec:taxonomy} introduces background information in terms of mobile operating systems, the basic notion of side-channel attacks, and related work. 
In Section~\ref{sec:new_categorization_system}, we discuss different types of information leaks and provide a definition for software-only side-channel attacks. 
Furthermore, we introduce our new categorization system for modern side-channel attacks. We survey existing attacks in Sections~\ref{sec:local}, \ref{sec:vicinity}, and~\ref{sec:remote}, and we classify existing attacks according to our newly introduced classification system in Section~\ref{sec:big_picture}. 
We discuss existing countermeasures in Section~\ref{sec:countermeasures}. 
Finally, we discuss open issues, challenges, and future research directions in Section~\ref{sec:future_research} and conclude in Section~\ref{sec:conclusion}.

\section{Background} 
 \label{sec:taxonomy}
In this section, we introduce the basics of mobile security, define the general notion of side-channel attacks, and we establish the boundaries between side-channel attacks and other attacks on mobile devices. We stress that side-channel attacks do not exploit specific software vulnerabilities of the OS or any specific library, but instead exploit available information that either leaks unintentionally or that is (in some cases) published for benign reasons in order to infer sensitive information indirectly. 
Finally, we also discuss related work. 

\subsection{A Primer on Smartphone Security}
Mobile devices, such as tablet computers and smartphones, are powerful multi-purpose computing platforms that enable many different application scenarios. Third-party applications can be easily installed in order to extend the basic functionality of these devices. Examples include gaming applications that make use of the many different sensors, office applications, banking applications, and many more. These examples clearly demonstrate that mobile devices are already tightly integrated into our everyday lives, which leads to sensitive data and information being stored and processed on these devices. 

In order to protect this information properly, modern mobile operating systems rely on two fundamental security concepts, \ie, the concept of application sandboxing and the concept of permission systems. For instance, on Android the underlying Linux kernel ensures the concept of sandboxed applications. 
Each application is assigned a user ID (UID), which allows the kernel to prevent applications from accessing resources of other applications. The permission system on the other hand allows applications to request access to specific resources outside of its sandbox, which typically includes resources that are considered as being sensitive or privacy relevant. Android also categorizes permissions depending on so-called protection levels. The two important categories of Android permissions are \emph{normal} permissions and \emph{dangerous} permissions, respectively. While \emph{normal} permissions are granted automatically during the installation procedure, \emph{dangerous} permissions must be explicitly granted by the user. Other mobile operating systems such as Apple's iOS rely on similar protection mechanisms. 

Besides these basic security concepts on the OS level, applications themselves rely on cryptographic primitives, cryptographic protocols, and dedicated security mechanisms to protect sensitive resources. 
For instance, applications rely on encryption primitives to protect sensitive information being stored on the device or when transmitting data over the Internet. Another example of a dedicated security mechanism is a personal identification number (PIN) required to access a specific service such as a banking application.

\subsection{Side-Channel Attacks}
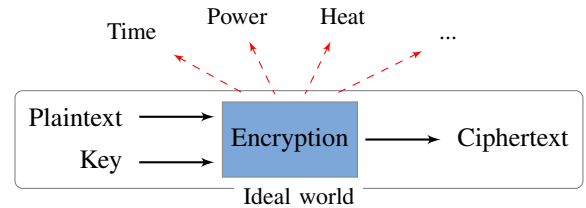
\begin{figure}
\tikzstyle{dev}=[rectangle, minimum width=1.5cm, minimum height=1cm,draw=gray!90,fill=RoyalBlue!50]
\begin{center}
\begin{tikzpicture}[auto, outer sep=3pt, node distance=5mm and 10mm,>=latex']
  \node[dev] (A) {Encryption};
  \coordinate (A input 1) at ($(A.north west)!0.25!(A.south west)$);
  \coordinate (A input 2) at ($(A.north west)!0.75!(A.south west)$);
  \coordinate (A output 1) at ($(A.north east)!0.5!(A.south east)$);

  \node[left=1cm] at (A input 1) (b) {Plaintext};
  \node[left=1cm] at (A input 2) (c) {Key};
  
  \draw[->,thick] (b) -- (A input 1);
  \draw[->,thick] (c) -- (A input 2);

  \node[right=1cm] at (A output 1) (d) {Ciphertext};
  \draw[<-,thick] (d) -- (A output 1);
  
  \node[draw=gray,rectangle,name=box, rounded corners=.1cm, inner sep=10pt, inner ysep=15pt, minimum width=7.55cm, minimum height=1.3cm] at (.12,0) {};
  \node[fill=white, text=black,font=\small] at (box.south) {Ideal world};
 
  \foreach \x in {1,2,3,4} {
    \coordinate (A leak \x) at ($(A.north west)!\x/5!(A.north east)$);
  }
  \node[above=.5cm, xshift=-1.5cm] at (A leak 1) (1) {\small Time}; 
  \draw[<-,dashed,red] (1) -- (A leak 1);
  
  \node[above=.7cm, xshift=-.5cm] at (A leak 2) (2) {\small Power}; 
  \draw[<-,dashed,red] (2) -- (A leak 2);
 
  \node[above=.7cm, xshift=.5cm] at (A leak 3) (3) {\small Heat}; 
  \draw[<-,dashed,red] (3) -- (A leak 3);
  
  \node[above=.5cm, xshift=1.5cm] at (A leak 4) (4) {\small ...}; 
  \draw[<-,dashed,red] (4) -- (A leak 4);

\end{tikzpicture}
\end{center}
\caption{An implementation produces unintended output as a byproduct.} 
\label{fig:real}
\end{figure}

Although the above mentioned concepts are secure (or are typically considered as being secure) in theory, a specific implementation of such a mechanism is not necessarily secure in practice. Since side-channel attacks have been extensively used to attack cryptographic implementations, let us consider the following illustrative example. In an ideal world, an implementation of a cryptographic algorithm takes a specific input and produces a specific (intended) output. 
For example, an encryption algorithm takes the plaintext as well as cryptographic key material to produce the ciphertext. 
However, in practice, an implementation of an encryption algorithm usually also ``outputs'' unintended information as a byproduct of the actual computations. Such unintended information leakage might be a different power consumption or a different execution time due to instructions being conditionally executed depending on the processed data (cf. Figure~\ref{fig:real}). 
Attacks exploiting such unintended information leaks are denoted as side-channel attacks and have been impressively used to bypass or break protection mechanisms such as encryption algorithms. 

Subsequently, we discuss the general notion of side-channel attacks. 
We distinguish between passive side-channel attacks, as in the example above, and active side-channel attacks. 

\textbf{Passive Side-Channel Attacks.} The general notion of a passive side-channel attack can be described by means of three main components, \ie, \emph{target}, \emph{side channel}, and \emph{attacker}. 
A \emph{target} represents anything of interest to possible attackers. During the computation or operation of the target, it influences a \emph{side channel} (physical or logical properties) and thereby emits potential sensitive information. 
An \emph{attacker} who is able to observe these side channels potentially learns useful information related to the actual computations or operations performed by the target. Therefore, an attacker models possible effects of specific causes. Later on, careful investigations of observed effects can then be used to learn information about possible causes. 

\textbf{Active Side-Channel Attacks.} An active attacker tampers with the device or modifies/influences the targeted device via a side channel, \eg, via an external interface or environmental conditions. Thereby, the attacker influences the computation/operation performed by the device in a way that allows to bypass specific security mechanisms directly or that leads to malfunctioning, which in turn enables possible attacks, \eg, indirectly via the leaking side-channel information or directly via the (erroneous) output of the targeted device. 

\begin{figure}
\tikzset{vertex style/.style={
    draw=#1,
    thick,
    color=black,
    fill=#1!50, 
    text=black,
    circle,
    minimum width=4.9cm, 
    minimum height=1cm,
    font=\Large,
    outer sep=3pt,
  },
  text style/.style={
    text=black,
    font=\large,
    above
  },
  example style/.style={
    draw=#1,
    thick,
    fill=#1!70,
    text=white, 
    circle,
    font=\footnotesize
  }
}
\centering
\resizebox{.99\columnwidth}{!}{
 \begin{tikzpicture}[node distance=2.75cm,>=stealth']
 \node[align=center,vertex style=RoyalBlue] (T) {\textbf{Target}\\(\eg, crypto,\\keyboard,\\behavior)};
 \node[align=center, vertex style=RoyalBlue, right of=T, xshift=10em] (SCS) {\textbf{Side channel}\\ (\ie, physical or\\logical properties)} edge [<-,thick,black,bend right=35] node[text style, yshift=.75em]{(1) influences} (T)
  (T) edge [<-,thick,dashed,black, bend right=35] node[text style, below, yshift=-1.25em]{(4) influences} (SCS);
 
 \node[align=center,vertex style=RoyalBlue, right of=SCS, xshift=10em] (O) {\textbf{Attacker}\\ (\eg, device, chip, \\wire, software)}
   edge [<-,thick,black,bend right=35] node[text style,yshift=.75em]{(2) is observed} (SCS)
   (SCS) edge [<-,thick,black, dashed, bend right=35] node[text style, below, yshift=-1.25em]{(3) modifies/influences} (O);
\end{tikzpicture}
}
\caption{General notion of passive ($\longrightarrow$) and active ($\dashleftarrow$) side-channel attacks.}
\label{fig:sca_notion}
\end{figure}
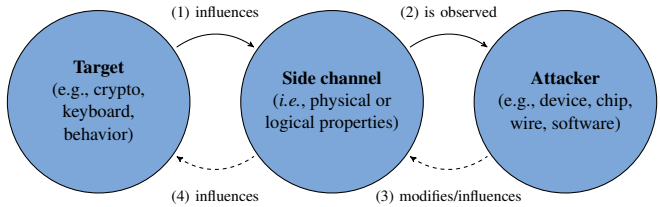

Figure~\ref{fig:sca_notion} depicts the general notion of side-channel attacks. A target emits sensitive information as it influences specific side channels. 
For example, physically operating a smartphone via the touchscreen,~\ie, the touchscreen input represents the target, causes the smartphone to undergo specific movements and accelerations in all three dimensions. 
In this case, one possible side channel is the acceleration of the device (a physical property), which can be observed via the embedded accelerometer sensor and accessed by an app via the official Sensor API. 

%
%

The relations defined via the solid arrows, \ie, \emph{target} $\longrightarrow$ \emph{side channel} $\longrightarrow$ \emph{attacker}, represent passive side-channel attacks. 
The relations defined via the dashed arrows, \ie, \emph{target} $\dashleftarrow$ \emph{side channel} $\dashleftarrow$ \emph{attacker}, represent active side-channel attacks where the attacker actively influences/manipulates the target via a side channel. 
Thereby, the attacker either tries (i) to enforce behavior that allows to bypass security mechanisms directly, or (ii) to observe leaking side-channel information or the (sometimes erroneous) output of the targeted device. 
Hence, a passive side-channel attack consists of steps (1) and (2), whereas an active side-channel attack also includes steps (3) and (4).

\textbf{Differentiation From Other Attacks.} 
Irrespective of whether an attacker is passive or active, we only consider side-channel attacks. 
Side-channel attacks \emph{do not} exploit software bugs or anomalies within the OS or apps that, for example, allow to access the main communication channel directly. 
For example, buffer overflow attacks allow to access the main communication channel directly (\ie, the main memory) and, thus, do not represent side-channel attacks. 
 
Similarly, we also do not consider other attacks that learn information that is available from the main channel. 
For example, Luzio~\etal~\cite{DBLP:conf/infocom/LuzioMS16} exploited \mbox{Wi-Fi} probe-requests, which contain the service set identifier (SSID) of preferred Wi-Fi hotspots in clear. 
These probe-requests allow mobile devices to determine nearby \mbox{Wi-Fi} hotspots in order to preferably connect to already known hotspots. 
These attacks do not represent side-channel attacks as the learned information is directly available from the main channel. 

Furthermore, we also do not survey covert channels where two entities (\eg, processes) communicate over a channel that is not explicitly provided by the platform or the operating system. 
Although identified side channels can in general also be used as a covert channel, \ie, as a means to stealthily communicate between two processes whereby one process influences the side channel and the other one observes it, we do not explicitly survey covert channels such as~\cite{SpolaorACNS2017} in this paper. 
Nevertheless, our newly introduced classification system can also be used to classify covert channels.

\begingroup
\xpatchcmd\rxmark{\xmark}{ }{}{}
\setlength{\aboverulesep}{0pt}
\setlength{\belowrulesep}{0pt}
\begin{table*}
\centering
\caption{Existing surveys and what they focus on. Upper part: surveys on mobile security. Lower part: surveys on side-channel attacks.} 
\label{tab:comparison_of_surveys}
\rowcolors{2}{white}{lightgray!40}
\begin{tabular}{llcc}
\toprule
\multirow{1}{*}{\rule{0pt}{2ex}\textbf{Year}}                         &\multirow{1}{*}{\rule{0pt}{2ex}\textbf{Survey}}                         & \multirow{1}{*}{\rule{0pt}{2ex}\textbf{Platform}} & \multirow{1}{*}{\rule{0pt}{2ex}\textbf{Topic}} \\
\midrule
2011 & Enck~\cite{DBLP:conf/iciss/Enck11} 					& Smartphone & Malware/app analysis and protection mechanisms  \\
2013 & La Polla~\etal~\cite{DBLP:journals/comsur/PollaMS13}                     & Smartphone & Threats and vulnerabilities, focusing on the period 2004--2011	\\
2014 & Suarez-Tangil~\etal~\cite{DBLP:journals/comsur/Suarez-TangilTPR14} 	& Smartphone & Threats and vulnerabilities, focusing on the period 2010--2013 \\ 
2015 & Faruki~\etal~\cite{DBLP:journals/comsur/FarukiBLGGCR15}           	& Smartphone & Threats and vulnerabilities, focusing on the period 2010--2014 \\ 
2015 & Rashidi and Fung~\cite{DBLP:journals/jowua/RashidiF15}           	& Smartphone & Analysis techniques to cope with malware  \\ 
2016 & Sadeghi~\etal~\cite{7583740}                                     	& Smartphone & Tools and techniques to identify malware \\ 
2017 & Tam~\etal~\cite{DBLP:journals/csur/TamFASC17}                            & Smartphone & Analysis techniques to identify malware\\ 
\midrule
2014 & Tunstall~\cite{Tunstall2017}		                                & Smart card &  Side-channel attacks on cryptographic algorithms\\
2007 & Zander~\etal~\cite{DBLP:journals/comsur/ZanderAB07} 			& PC         & Covert channels via computer network protocols \\
2017 & Biswas~\etal~\cite{DBLP:journals/csur/BiswasGN17}			& PC	     &   Timing channels, focusing on microarchitectural attacks          \\
2016 & Ge~\etal~\cite{Ge2016}                                            	& Cloud      & Microarchitectural attacks  \\
2016 & Szefer~\cite{DBLP:journals/iacr/Szefer16}                       		& Cloud      & Microarchitectural attacks \\
2017 & Ullrich~\etal~\cite{DBLP:journals/comsur/UllrichZFW17}		        & Cloud      &    Network-based side channels (and communication channels) \\
2017 & Betz~\etal~\cite{DBLP:journals/ett/BetzWM17}	                        & Cloud	     &  Communication channels        \\
2016 & Xu~\etal~\cite{DBLP:journals/csur/XuSJSLZDJLQLK16}                       & Smartphone 	&  Attacks \& defense measures\\ 
2016 & Hussain~\etal~\cite{DBLP:journals/percom/HussainAZZKAA16}                &  Smartphone 	& Sensor-based keylogging attacks \\
2016 & Nahapetian~\cite{DBLP:conf/ccnc/Nahapetian16}                            & Smartphone 	& Sensor-based keylogging attacks \\
\bottomrule
\end{tabular} 
\end{table*}
\endgroup

\subsection{Related Surveys}
 \label{sec:related_surveys}
In this section, we discuss surveys on mobile security, as well as side-channel attacks on smart cards, PCs, cloud infrastructures, and smartphones. 

\textbf{Surveys on Mobile Security.}
Most surveys on mobile security primarily focused on malware in general, and many of these surveys only mention side-channel attacks as a side node. 
Enck~\cite{DBLP:conf/iciss/Enck11} surveyed possible protection mechanisms beyond the standard protection mechanisms provided by Android.
These include tools that analyze permissions and action strings (within the Android Manifest) to assess the risk of Android apps, policy-based approaches that allow a more fine-grained protection of Android apps, as well as static and dynamic code analysis tools to perform application analysis, which in turn allows to detect malware.

La Polla~\etal~\cite{DBLP:journals/comsur/PollaMS13} surveyed threats and vulnerabilities (\ie, botnets, Trojans, viruses, and worms) with a focus on work published from 2004 until 2011. 
Suarez-Tangil~\etal~\cite{DBLP:journals/comsur/Suarez-TangilTPR14} and Faruki~\etal~\cite{DBLP:journals/comsur/FarukiBLGGCR15} continued this line of research for the period from 2010 until 2013, and from 2010 until 2014, respectively.
 
Rashidi and Fung~\cite{DBLP:journals/jowua/RashidiF15} surveyed techniques (\eg, based on static and dynamic code analysis) to cope with malware on mobile devices
and Sadeghi~\etal~\cite{7583740} surveyed tools and analysis techniques to identify malware. 
In addition, Sadeghi~\etal~provided a ``survey of surveys'' discussing surveys and their main contributions in more detail. 
We refer to their work for a more detailed investigation of malware analysis techniques and further literature on this topic.
Tam~\etal~\cite{DBLP:journals/csur/TamFASC17} surveyed mobile malware analysis techniques (static, dynamic, hybrid) as well as malware tactics to hinder analysis (obfuscation).

\textbf{Surveys on Side-Channel Attacks.}
The survey of Tunstall~\cite{Tunstall2017} focused on smart card security, in particular side-channel attacks against cryptographic algorithms. 

Zander~\etal~\cite{DBLP:journals/comsur/ZanderAB07} surveyed covert channels via computer network protocols, and Biswas~\etal~\cite{DBLP:journals/csur/BiswasGN17} conducted an in-depth study on network timing channels (remote timing side channels) as well as in-system timing channels (focusing on hardware-based timing channels such as cache attacks) on commodity PCs. 
They surveyed timing channels according to their suitability for covert channels, timing side channels, and network flow watermarking (\eg, to de-anonymize Tor). 

Regarding cloud computing platforms, Ge~\etal~\cite{Ge2016} and Szefer~\cite{DBLP:journals/iacr/Szefer16} surveyed microarchitectural attacks with a focus on cache attacks. 
Ullrich~\etal~\cite{DBLP:journals/comsur/UllrichZFW17} focused on network-based covert channels and network-based side channels in cloud settings. 
Betz~\etal~\cite{DBLP:journals/ett/BetzWM17} focused on covert channels and mentioned a few side-channel attacks in the cloud setting. 

The focus of our paper is on side-channel attacks against mobile devices. 
Surveys about this topic are quite scarce and consider specific types of side-channel attacks only. 
Xu~\etal~\cite{DBLP:journals/csur/XuSJSLZDJLQLK16} surveyed attacks and defenses on Android at a broader scale and thereby provide a comprehensive overview of the research landscape. 
They considered system privilege escalation, issues in the permission model, side channels and covert channels (a high-level overview of exploits considering the accelerometer, the CPU cache, and the procfs), feature abuses, malware detection, and app repackaging. 
Hussain~\etal~\cite{DBLP:journals/percom/HussainAZZKAA16} and Nahapetian~\cite{DBLP:conf/ccnc/Nahapetian16} surveyed sensor-based keylogging attacks. 
However, a systematic survey and classification of all existing categories of side-channel attacks on mobile devices does not exist so far. 
Hence, we close this gap in this paper. 

Table~\ref{tab:comparison_of_surveys} summarizes the main focus of the above discussed surveys and provides references for the interested reader.

\section{Taxonomy of Side Channels}
 \label{sec:new_categorization_system}
In this section, we discuss the different types of information leaks, how the key enablers presented in Section~\ref{sec:intro} enable so-called \emph{software-only attacks} on today's smartphones, and the generic adversary model followed by software-only attacks. 
Finally, we present our new categorization system. 

\subsection{Types of Information Leaks}
\begin{figure}
\centering

 \tikzset{
  basic/.style  = {draw, text width=2cm, font=\sffamily, rectangle},
  root/.style   = {basic, rounded corners=2pt, thin, align=center,
                   fill=RoyalBlue!50},
  level 2/.style = {basic, rounded corners=6pt, thin,align=center, fill=RoyalBlue!40,
                   text width=12em, sibling distance=15mm},
  level 3/.style = {basic, thin, align=left, fill=RoyalBlue!10, text width=10em}
 }
 \resizebox{0.8\columnwidth}{!}{
\begin{tikzpicture}[
  level 1/.style={sibling distance=50mm},
  edge from parent/.style={->,draw},
  >=latex]
  
  \node[root,style={text width=8cm}] {Side-channel information leaks}
  child {node[level 2] (c1) {Unintended information leaks}}
  child {node[level 2] (c2) {Information published on purpose}};

  \begin{scope}[every node/.style={level 3}]
  \node [below of = c1] (c11) {Execution time};
  \node [below of = c11] (c12) {Power consumption};
  \node [below of = c12] (c13) {EM emanation};

  \node [below of = c2] (c21) {Memory footprint};
  \node [below of = c21] (c22) {Sensor information};
  \node [below of = c22] (c23) {Data consumption};
  \end{scope}
  \foreach \value in {1,2,3}{
    \draw[->] (c1.175) |- (c1\value.west);
    \draw[->] (c2.175) |- (c2\value.west);
  }
\end{tikzpicture}
}
\caption{Types of side-channel information leaks.}
\label{fig:information_leaks}
\end{figure}
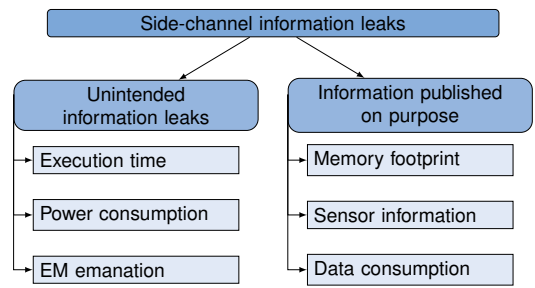

Considering side-channel attacks on mobile devices, we identify two types of information leaks, namely \emph{unintended information leaks} and \emph{information published on purpose}. Figure~\ref{fig:information_leaks} depicts these two types of information leaks. Informally, side-channel attacks exploiting unintended information leaks can be considered as ``traditional'' side-channel attacks since this category has already been extensively analyzed during the smart card era~\cite{DBLP:books/daglib/0017272}. 
For example, unintended information leaks include the execution time, the power consumption, or the electromagnetic emanation of a computing device. This type of information leak is considered as unintended because smart card designers and developers did not plan to leak the timing information or power consumption of computing devices on purpose.

The second category of information leaks (referred to as \emph{information published on purpose}) is mainly a result of the ever-increasing number of features provided by today's smartphones. 
In contrast to unintended information leaks, the exploited information is published on purpose and for benign reasons. 
For instance, specific features require the device to share (seemingly harmless) information and resources with apps running in parallel on the system. 
This information is either shared by the OS directly (\eg, via the procfs) or through the official Android API.\footnote{In the literature, some of the information leaks through the procfs are also denoted as \emph{storage side channels}~\cite{DBLP:conf/ccs/XiaoRZ15}.} 
Although this information is extensively used by many legitimate applications for benign purposes, it sometimes turns out to leak sensitive information and, thus, leads to devastating side-channel attacks. 

Many investigations impressively demonstrated that seemingly harmless information allows to infer sensitive information that is protected by dedicated security mechanisms, such as permissions. 
Examples of such seemingly harmless information are the memory footprint of an application as well as the data-usage statistics that keep track of the amount of incoming and outgoing network traffic. 
Both, the memory footprint~\cite{DBLP:conf/sp/JanaS12a} as well as the data-usage statistics~\cite{DBLP:conf/wisec/SpreitzerGKM16}, allow to infer a user's visited websites. 
The fundamental design weakness of assuming information as being innocuous (\eg, the memory footprint or the data-usage statistics) means that it is not protected by dedicated permissions. 

Furthermore, the second category seems to be more dangerous in the context of smartphones as new features are frequently added and new software interfaces allow to access an unlimited number of unprotected resources. Even developers taking care of secure implementations in the sense of unintended information leaks, \eg, by providing constant-time crypto implementations and taking care of possible software vulnerabilities such as buffer overflow attacks, inevitably leak sensitive information due to shared resources, the OS, or the Android API. Additionally, the provided software interfaces to access information and shared resources enable so-called \emph{software-only attacks}, \ie, side-channel attacks that only require the execution of software. This clearly represents an immense threat as these attacks (1) do not exploit any obvious software vulnerabilities,  (2) do not rely on specific privileges or permissions, and (3) can be conducted remotely via seemingly harmless apps or even websites.

\subsection{Software-only Side-Channel Attacks}
Irrespective of whether a physical property (\eg, execution time~\cite{DBLP:conf/uss/YaromF14} and power consumption~\cite{DBLP:conf/uss/MichalevskySVBN15}) or a logical property (\eg, memory footprint~\cite{DBLP:conf/sp/JanaS12a} and data-usage statistics~\cite{DBLP:conf/ccs/ZhouDHNPWGN13,DBLP:conf/wisec/SpreitzerGKM16}) are exploited, many of these information leaks can be exploited by means of \emph{software-only attacks}. 
More specifically, software-only attacks exploit leaking information without additional equipment, which was required for traditional side-channel attacks. 
For example, an oscilloscope is necessary to measure the power consumption of a smart card during its execution, or an EM probe is necessary to measure the EM emanation. 
In contrast, today's smartphones allow an impressive number of side-channel leaks to be exploited via software-only attacks. 
Besides, an attack scenario that requires the user to install an (unprivileged) application---\ie, an addictive game---is entirely reasonable in an appified ecosystem. 

For side-channel attacks in general, it does not matter whether the leaking information is collected via dedicated equipment or whether an unprivileged app collects the leaking information directly on the device under attack (software-only attacks). Interestingly, however, the immense amount of information published on purpose also allows to observe physical properties of the device as well as physical interactions with the device. 
Consequently, \emph{software-only side channel attacks} have gained increasing attention in the last few years and impressive attacks are being continuously published.

\textbf{Runtime-Information Gathering Attacks.} 
Zhang~\etal~\cite{DBLP:conf/sp/ZhangY0ZW15} coined the term runtime-information gathering (RIG) attack, which refers to attacks that require a malicious app to run side-by-side with a victim app on the same device in order to collect runtime information of the victim. 
According to Zhang~\etal~\cite[p.~1]{DBLP:conf/sp/ZhangY0ZW15} ``(RIG) here refers to any malicious activities that involve collecting the data produced or received by an app during its execution, in an attempt to directly steal or indirectly infer sensitive user information''. 
The crucial point in their definition is the distinction between \emph{directly stealing} and \emph{indirectly inferring} sensitive information. 
Inferring sensitive information indirectly is done by means of side-channel attacks. Hence, this generic class of attacks also includes a subset of side-channel attacks, especially side-channel attacks that can be launched via software-only attacks. However, RIG attacks also include attacks that we do not consider as side-channel attacks, \ie, attacks that \emph{directly} steal sensitive information. 
For example, RIG attacks also include attacks where apps request permissions which are exploited for (more obvious) attacks such as requesting the permission to access the microphone in order to eavesdrop on phone conversations. 

Screenmilker~\cite{DBLP:conf/ndss/LinLZW14}---an attack exploiting ADB\footnote{The Android Debug Bridge (ADB) is a command line tool that allows to execute privileged commands on devices where USB debugging is activated.} capabilities to take screenshots programmatically---is also considered being a RIG attack. We do not consider such attacks as side-channel attacks because these attacks exploit implementation flaws, \ie, the exploited screenshot tool does not implement any authentication mechanism and hence any application can take screenshots programmatically. Similarly, we do not consider buffer overflow attacks as side-channel attacks because buffer overflow attacks represent a software vulnerability and allow to access the main channel directly, for example, by reading the main memory directly. Side-channel attacks, however, attack targets that are secure from a software perspective and still leak information unintentionally. 

\begin{figure}[t]
 \centering
 \resizebox{.75\columnwidth}{!}{
 \begin{tikzpicture}
   \begin{scope}
   \def\firstellipse{(0,0) ellipse [x radius=1.5, y radius=2.5, rotate=90]}
   \def\firstcircle{(-2.5,0) circle (1.5)}
   \def\secondcircle{(+2.5,0) circle (1.5)}
   
   \draw \firstellipse;
   \draw \firstcircle;
   \draw \secondcircle;
   
    \begin{scope}
      \clip \firstcircle;
      \fill[NavyBlue!60!white] \firstellipse;
    \end{scope}
    \begin{scope}
      \clip \secondcircle;
      \fill[NavyBlue!60!white] \firstellipse;
    \end{scope}
   \end{scope}

   \node[align=center] at (0,0) {RIG\\attacks};
   \node[align=center] at (-3.2,0) {Physical\\attacks};
   \node[align=center,rotate=0] at (3.2,0) {Logical\\attacks};
   \draw[<-] (-1.75,0)
           -- (0,2) node [align=center,above] {SW-only side-channel attacks};
   \draw[<-] (1.75,0) -- (0,2);
 \end{tikzpicture}
 }
 \caption{SW-only side-channel attacks allow to exploit physical as well as logical properties.}
 \label{fig:sw_only}
\end{figure}
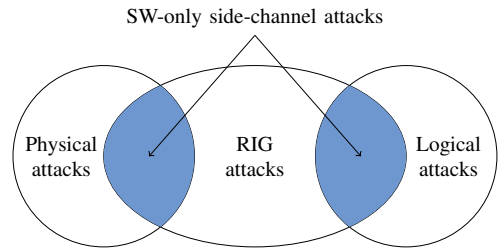

Figure~\ref{fig:sw_only} illustrates the new type of software-only side-channel attacks that allow to exploit both, physical properties as well as software features (logical properties), without additional equipment. 
Attacks exploiting information leaks resulting from hardware components, \eg, the power consumption, are classified as (physical) attacks exploiting physical properties. 
Attacks exploiting information leaks resulting from software components, \eg, statistics about network traffic, are classified as (logical) attacks exploiting logical properties.

As software-only attacks also rely on software being executed side-by-side with the victim application, software-only attacks are a sub-category of RIG attacks. It should be noted that physical attacks on smartphones might still rely on dedicated hardware and some logical attacks can also be conducted without running software on the device under attack. Such attacks are covered by the non-overlapping areas of ``physical attacks'' and ``logical attacks'' in Figure~\ref{fig:sw_only}. However, physical attacks that cannot be conducted by running software on the device are more targeted attacks as they require attackers to be in physical presence of the device. 

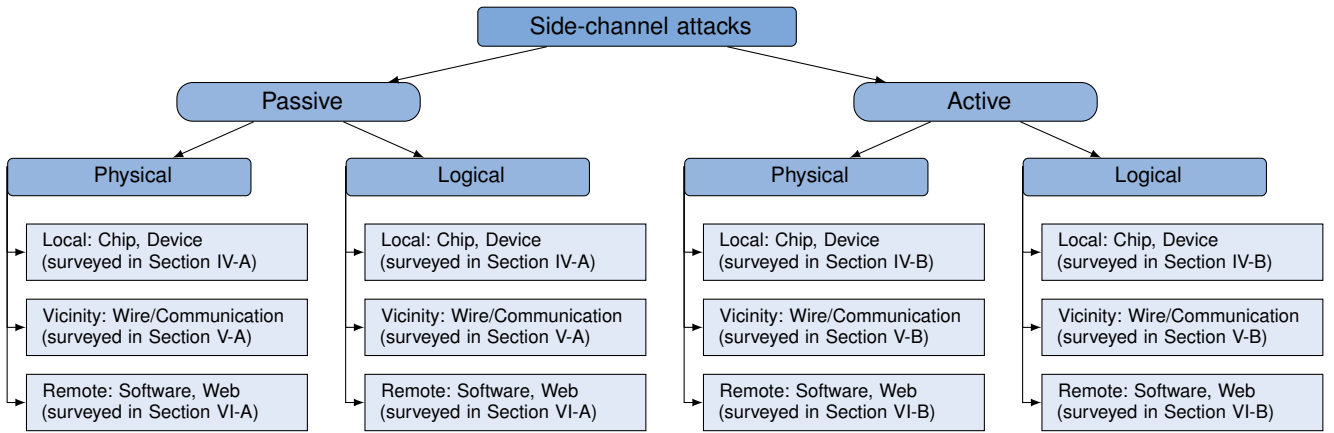
\begin{figure*}
\centering
\begin{tikzpicture}[
  root/.style      = {draw, rectangle, font={\sffamily,\small}, rounded corners=2pt, thin, align=center, fill=RoyalBlue!50, text width=4cm},
  level 1/.style   = {draw, rectangle, font={\sffamily,\small}, rounded corners=6pt, thin, align=center, fill=RoyalBlue!40, text width= 3cm, sibling distance=90mm, level distance=10mm},
  level 2/.style   = {draw, rectangle, font={\sffamily,\footnotesize}, rounded corners=2pt, thin, align=center, fill=RoyalBlue!40, text width= 3cm, sibling distance=45mm},
  level 3/.style   = {draw, rectangle, font={\sffamily,\scriptsize}, rounded corners=0pt, thin, align=left, fill=RoyalBlue!10, text width= 3.5cm, sibling distance=20mm},
  edge from parent/.style={->,draw,black},
  >=latex]

\node[root] {Side-channel attacks}
  child {node[level 1] (c1) {Passive}
         child {node[level 2] (c11) {Physical}
               }
         child {node[level 2] (c12) {Logical}}
        }
  child {node[level 1] (c2) {Active}
	 child {node[level 2] (c21) {Physical}}
         child {node[level 2] (c22) {Logical}}
        };

\begin{scope}[every node/.style={level 3}]
\node [below of = c11, xshift=.5cm] (c111) {Local: Chip, Device\\~(surveyed in Section~\ref{subsec:local_passive})};
\node [below of = c111] (c112) {Vicinity: Wire/Communication\\~(surveyed in Section~\ref{subsec:vicinity_passive})};
\node [below of = c112] (c113) {Remote: Software, Web\\~(surveyed in Section~\ref{subsec:remote_passive})};

\node [below of = c12, xshift=.5cm] (c121) {Local: Chip, Device\\~(surveyed in Section~\ref{subsec:local_passive})};
\node [below of = c121] (c122) {Vicinity: Wire/Communication\\~(surveyed in Section~\ref{subsec:vicinity_passive})};
\node [below of = c122] (c123) {Remote: Software, Web\\~(surveyed in Section~\ref{subsec:remote_passive})};

\node [below of = c21, xshift=.5cm] (c211) {Local: Chip, Device\\~(surveyed in Section~\ref{subsec:local_active})};
\node [below of = c211] (c212) {Vicinity: Wire/Communication\\~(surveyed in Section~\ref{subsec:vicinity_active})};
\node [below of = c212] (c213) {Remote: Software, Web\\~(surveyed in Section~\ref{subsec:remote_active})};

\node [below of = c22, xshift=.5cm] (c221) {Local: Chip, Device\\~(surveyed in Section~\ref{subsec:local_active})};
\node [below of = c221] (c222) {Vicinity: Wire/Communication\\~(surveyed in Section~\ref{subsec:vicinity_active})};
\node [below of = c222] (c223) {Remote: Software, Web\\~(surveyed in Section~\ref{subsec:remote_active})};
\end{scope}
 \foreach \value in {1,2,3}{
  \draw[->] (c11.175) |- (c11\value.west);
  \draw[->] (c12.175) |- (c12\value.west);
  \draw[->] (c21.175) |- (c21\value.west);
  \draw[->] (c22.175) |- (c22\value.west);
 }
 \end{tikzpicture}
\caption{Proposed classification system for side-channel attacks: (1) passive vs active, (2) physical properties vs logical properties, (3) local attackers vs vicinity attackers vs remote attackers.}
\label{fig:basic_classification}
\end{figure*}

\subsection{Adversary Model and Attack Scenario}
In contrast to traditional attacks that require an attacker to have the device under physical control or to be physically present with the victim, the adversary model for most (existing) side-channel attacks on smartphones shifted the scope to remote software execution by means of apps or websites. This also increases the scale of these attacks. While traditional side-channel attacks targeted only a few devices, modern side-channel attacks target possibly millions of devices or users at the same time.  
With this general overview of the adversary model in mind, most software-only attacks usually consider the following two-phase attack scenario for passive attacks. 

\textbf{Training Phase.} 
In the training phase, the attacker ``profiles'' actions or events of interest, either during an online phase on the attacked device or during an offline phase in dedicated environments. Sometimes the training phase includes the training of a machine-learning model such as a supervised classifier. More abstractly, the attacker builds ``templates'' based on events of interest. In addition, the attacker crafts an app (or website) that ideally does not require any permissions or privileges in order to avoid raising the user's suspicion. This app is used in the attack phase to gather leaking information. 

\textbf{Attack Phase.}
The attack phase usually consists of three steps. (1) A malicious application---that is hidden inside a popular app---is spread via existing app markets. After installation, this malicious app waits in the background until the targeted app/action/event starts and then (2) it observes the leaking side-channel information. Based on the gathered information, (3) it employs the previously established model or templates to infer secret information. Depending on the complexity of the inference mechanism, \eg, the complexity of the machine-learning classifier, the gathered side-channel information could also be sent to a remote server, which then performs the heavy computations to infer the secret information. 

\subsection{A New Categorization System}
Based on our observations we propose a new categorization system as depicted in Figure~\ref{fig:basic_classification}. 
More specifically, we classify side-channel attacks along three axes.
\begin{enumerate}
 \item \emph{Passive} vs \emph{active}: This category distinguishes between attackers who passively observe leaking side-channel information and attackers who also actively influence the target via any side channel. For instance, an attacker can manipulate the target, its input, or its environment via any side channel in order to subsequently observe leaking information via abnormal behavior of the target (cf.~\cite{DBLP:books/daglib/0017272}) or to bypass security mechanisms. 
 
 \item \emph{Physical properties} vs \emph{logical properties}: This category classifies side-channel attacks according to the exploited information, \ie, depending on whether the attack exploits physical properties (hardware) or logical properties (software features). 
 Physical properties include the power consumption, the electromagnetic emanation, or the physical movements of a smartphone during the operation. 
 Logical properties include usage statistics provided by the operating system, such as the data-usage statistics or the memory footprint of an application. 
 
 \item \emph{Local attackers} vs \emph{vicinity attackers} vs \emph{remote attackers}: Side-channel attacks are classified depending on whether or not the attacker must be in physical proximity/vicinity of the target. \emph{Local attackers} clearly must be in (temporary) possession of the device or at least in close proximity. Depending on whether the adversary also needs to remove the package in order to access the chip, we classify local attackers into attackers that need access to the chip or only the device itself. 
 \emph{Vicinity attackers} are able to wiretap or eavesdrop the network communication of the target or to be 
 somewhere in the vicinity of the target. 
 \emph{Remote attackers} only rely on software execution on the targeted device, \eg, either by means of executing software on the targeted device or by means of websites. 
 Clearly, the scale increases significantly for these three attackers as a local attacker relies on stronger assumptions than a remote attacker. Especially the immense number of software-only attacks (that allow to conduct side-channel attacks remotely) stress the need for this category. 
\end{enumerate}

Subsequently, we briefly survey existing attacks according to our new classification system. 
Although the focus of this paper is on side-channel attacks against mobile devices, we also discuss attacks that have been applied in the smart card or desktop/cloud setting, as today's smartphones are vulnerable to (all or most of the) existing side-channel attacks against these platforms as well. 
As mentioned before, we do not explicitly focus on Android devices, but the majority of existing papers investigate side-channel attacks on Android. 

We start with \emph{local} side-channel attacks in Section~\ref{sec:local}, continue with \emph{vicinity} side-channel attacks in Section~\ref{sec:vicinity}, and finally we discuss \emph{remote} side-channel attacks in Section~\ref{sec:remote}. Each of these sections is further divided into \emph{passive attacks} and \emph{active attacks}. Note that this structure reflects our proposed classification system. However, for the sake of readability the structure of the subsections does not reflect the categorization of physical properties and logical properties.

\section{Local Side-Channel Attacks}
 \label{sec:local}
In this section, we survey side-channel attacks that require a local adversary. 
Some of these attacks will show that the transition between local attacks and vicinity attacks is seamless as the distance between the victim (device) and the attacker can be increased, especially in case of some passive attacks. 

\subsection{Passive Attacks}
 \label{subsec:local_passive}
We start with traditional side-channel attacks that aim to break insecure cryptographic implementations (of mathematically secure primitives). Besides, we discuss attacks that target the user's interaction with the device as well as the user's input on the touchscreen, \ie, attacks that result from the inherent nature of mobile devices.

\attackSection{Power Analysis Attacks}
The actual power consumption of a computing device or implementation depends on the processed data and the executed instructions. Power analysis attacks exploit this information leak to infer sensitive information. As the name suggests, the power consumption, typically measured as the voltage drop across a resistor inserted in the supply line, serves as the side channel. State-of-the-art printed circuit board designs (PCB-designs), including multi-layer routing as well as surface mounted devices (SMD), and packaging techniques (\eg, ball-grid array) make it hard to access the appropriate power supply lines in modern smartphones without permanent modifications. Therefore, in contrast to smart cards, measuring the power consumption became less relevant for side-channel attacks targeting smartphones. 

Depending on whether a single measurement trace or multiple traces are required, we distinguish between simple power analysis (SPA) attacks and differential power analysis (DPA) attacks, as defined by Kocher~\etal~\cite{DBLP:conf/crypto/KocherJJ99}. SPA attacks rely on the interpretation of power traces in order to reveal, for example, the sequence of executed instructions, which allows to break implementations where the executed instructions depend on secret data. 
However, the power consumption also depends on the processed data, although the variations are smaller. Therefore, DPA attacks rely on statistical investigations of multiple traces in order to infer information about the processed data. 

\attacks
Messerges and Dabbish~\cite{DBLP:conf/smartcard/MessergesD99} exploited the power consumption of a smart card to attack the Data Encryption Standard (DES) algorithm. Hardly any side-channel attacks using a similar setup for measuring the power consumption targeting smartphones are published. 
Nevertheless, a coarse-grained power-consumption monitoring of smartphones allows to identify running apps, as demonstrated by Yan~\etal~\cite{DBLP:conf/internetware/YanGCM15}.

\attackSection{Electromagnetic Analysis Attacks}
Another way to attack the leaking power consumption of computing devices is to exploit electromagnetic emanations, which are usually easier to obtain since the power line cannot be accessed directly in general. 
Irrespective of whether the power trace is obtained directly via the power line or via electromagnetic emanations, these attacks are usually denoted as differential power analysis attacks.
In this context it is also worth to mention that depending on the used equipment (EM probes for capturing the electromagnetic emanation), targeting a specific location above the chip can improve the signal-to-noise ratio of the measurements. As a result of taking advantage of spatial information, the number of required measurements for a successful attack can be decreased. 

\attacks 
Traditional side-channel attacks exploiting the electromagnetic emanations of smart cards have also been applied on mobile devices. 
Gebotys~\etal~\cite{DBLP:conf/ches/GebotysHT05} demonstrated attacks on software implementations of the Advanced Encryption Standard (AES) and Elliptic Curve Cryptography (ECC) on Java-based PDAs. Later on, Nakano~\etal~\cite{DBLP:conf/wistp/NakanoSNSDGKM14} attacked ECC and RSA implementations of the default crypto provider (JCE) on Android smartphones, Goller and Sigl~\cite{DBLP:conf/cosade/GollerS15} attacked RSA implementations on Android, and Belgarric~\etal~\cite{DBLP:conf/ctrsa/BelgarricFMT16} attacked the Elliptic Curve Digital Signature Algorithm (ECDSA) implementation of Android's Bouncy Castle. In a similar manner, Genkin~\etal~\cite{DBLP:conf/ccs/GenkinPPTY16} attacked the OpenSSL implementation of ECDSA on Android and the CommonCrypto implementation of ECDSA on iOS, respectively. 

\attackSection{Differential Computation Analysis}
The basic idea of white-box crypto implementations is to embed the secret key into the software implementation in a way that prevents an attacker from extracting the key, even in case the adversary has access to the source code itself. Therefore, the key and the algorithm itself are merged such that the key is hidden inside the code and cannot be easily separated. The white-box attack model assumes that the adversary has full control over the device and the execution environment.  

\attacks
Bos~\etal~\cite{DBLP:conf/ches/BosHMT16} showed that binary instrumentation can be used to observe and control the intermediate state of white-box crypto implementations. Thereby, the instrumentation allows to precisely monitor the execution of the program and the observation of, \eg, the intermediate state and read/write accesses to memory, allow to profile program behavior. Based on the similarity to DPA attacks, Bos~\etal~denoted these attacks as differential computation analysis (DCA) attacks. Nevertheless, in contrast to DPA attacks, DCA attacks do not need to deal with any measurement noise. 

Although attacks against white-box crypto implementations have not been applied on mobile devices so far, such an attack scenario works for these devices as well. 

\attackSection{Smudge Attacks}
The most common input method on mobile devices is the touchscreen, \ie, users tap and swipe on the screen with their fingers. Due to the inherent nature of touchscreens, users always leave residues in the form of fingerprints and smudges on the screen. 

\attacks
Aviv~\etal~\cite{DBLP:conf/woot/AvivGMBS10} pointed out that side-channel attacks can be launched due to specific interactions with the smartphone or touchscreen-based devices in general. More specifically, forensic investigations of smudges (oily residues from the user's fingers) on the touchscreen allow to infer unlock patterns. Even after cleaning the phone or placing the phone into the pocket, smudges seem to remain most of the time. Hence, smudges are quite persistent which increases the threat of smudge attacks. Follow-up work considering an attacker who employs fingerprint powder to infer keypad inputs has been presented by Zhang~\etal~\cite{DBLP:conf/ccs/ZhangXLLLF12} and also an investigation of the heat traces---left on the screen due to finger touches---by means of thermal cameras has been performed~\cite{DBLP:conf/wisec/AndriotisTOY13}.

\attackSection{Shoulder Surfing and Reflections} 
Touchscreens of mobile devices optically/visually emanate the displayed content. Often these visual emanations are reflected by objects in the environment, such as sunglasses and tea pots~\cite{DBLP:conf/sp/BackesDU08,DBLP:conf/sp/BackesCDLW09}.

\attacks
Maggi~\etal~\cite{DBLP:conf/IEEEias/MaggiGB11} observed that touchscreen input can be recovered by monitoring the visual feedback (pop-up characters) on soft keyboards during the user input. Therefore, they rely on cameras that are pointed directly on the targeted screen. Raguram~\etal~\cite{DBLP:conf/ccs/RaguramWGMF11,DBLP:journals/tdsc/Raguram0XFGM13} observed that reflections, \eg, on the user's sunglasses, can also be used to recover input typed on touchscreens. However, the attacker needs to point the camera, used to capture the reflections, directly on the targeted user. Subsequently, they rely on computer vision techniques and machine learning techniques to infer the user input from the captured video stream. Xu~\etal~\cite{DBLP:conf/ccs/XuH0MF13} extended the range of reflection-based attacks by considering reflections of reflections. Although, they do not rely on the visual feedback of the soft keyboard but instead track the user's fingers on the smartphone while interacting with the device. 

By increasing the distance between the attacker and the victim, \eg, by relying on more expensive and sophisticated cameras, some of these attacks might as well be considered as vicinity attacks.

\attackSection{Hand/Device Movements}
Many input methods on various devices rely on the user operating the device with her hands and fingers. For instance, users tend to hold the device in their hands while operating it with their fingers. 

\attacks
Similar to reflections, Shukla~\etal~\cite{DBLP:conf/ccs/ShuklaKSP14} proposed to monitor hand movements as well as finger movements---without directly pointing the camera at the targeted screen---in order to infer entered PIN inputs. Sun~\etal~\cite{DBLP:conf/ndss/SunJCZZ016} monitored the backside of tablets during user input and detected subtle motions that can be used to infer keystrokes, while Yue~\etal~\cite{DBLP:conf/ccs/YueLFLRZ14} proposed an attack where the input on touch-enabled devices can be estimated from a video of a victim tapping on a touch screen. 

Again, by increasing the distance between the attacker and the victim, these attacks might also be considered as vicinity attacks, which demonstrates the seamless transition from local attacks to vicinity attacks for these types of attacks.

\subsection{Active Attacks}
 \label{subsec:local_active}
An active attacker also manipulates the target, its input, or its environment in order to subsequently observe leaking information via abnormal behavior of the target or to bypass security mechanisms directly. 
While the transition between local and vicinity attackers is seamless in case of passive attacks, active attacks always assume that the attacker is in possession of the device (at least temporary). 

Active attacks against cryptographic implementations date back to the works of Boneh~\etal~\cite{DBLP:conf/eurocrypt/BonehDL97} (a.k.a. Bellcore attack) who attacked RSA crypto systems, especially implementations based on the Chinese Remainder Theorem (CRT), by relying on random hardware faults that result in the output of an erroneous signature. 
Later, Biham and Shamir~\cite{DBLP:conf/crypto/BihamS97} coined the term differential fault analysis (DFA) attacks and demonstrated that the introduction of faults and observing differences in the output ciphertext allow to recover the secret key of symmetric primitives. 
The basic idea of these attacks is to solve algebraic equations based on erroneous outputs (and valid outputs). 

\attackSection{Clock/Power Glitching} 
Variations of the clock signal, \eg, overclocking, have been shown to be an effective method for fault injection on embedded devices in the past. One prerequisite for this attack is an external clock source. Microcontrollers applied in smartphones typically have an internal clock generator, making clock tampering difficult. Besides clock tampering, intended variations of the power supply represent an additional method for fault injection. With minor hardware modifications, power-supply tampering can be applied on most microcontroller platforms. 

\attacks
In \cite{ChipWhispererFIRaspberryPI} it is shown how to disturb the program execution of an ARM CPU on a Raspberry PI by underpowering, \ie, the supply voltage is set to ground (GND) for a short time. 
Due to the relatively easy application on modern microcontrollers, voltage-glitching attacks pose a serious threat for smartphones if attackers have physical access to the device. For instance, O'Flynn~\cite{DBLP:journals/iacr/OFlynn16} demonstrated that by shorting the power supply of an off-the-shelf Android smartphone, a fault can be introduced that leads to an incorrect loop count. 

\attackSection{Electromagnetic Fault Injection (EMFI)} 
Transistors placed on microchips can be influenced by electromagnetic emanation. EMFI attacks take advantage of this fact. These attacks use short (in the range of nanoseconds), high-energy EM pulses to, \eg, change the state of memory cells, resulting in erroneous calculations. In contrast to voltage glitching, where the injected fault is typically global, EMFI allows to target specific regions of a microchip by precisely placing the EM probe, \eg, on the instruction memory, the data memory, or CPU registers. Compared to optical fault injection, EMFI attacks do not necessarily require a decapsulation of the chip, which makes them more practical.

\attacks
Ordas~\etal~\cite{Ordas2016} reported successful EMFI attacks targeting the AES hardware module of a 32\,bit ARM processor. Rivi{\`{e}}re~\etal~\cite{DBLP:conf/host/RiviereNRDBS15} used EMFI attacks to force instruction skips and instruction replacements on modern ARM microcontollers. Considering the fact that ARM processors are applied in modern smartphones, EMFI attacks represent a serious threat for such devices.

\setlength{\aboverulesep}{0pt}
\setlength{\belowrulesep}{0pt}
\begin{table*}
\centering
\caption{Overview of local side-channel attacks and corresponding targets. \gcmark\ and \rxmark\ indicate whether or not a specific attack has been performed on the corresponding target.}
\label{tab:overview_local_attacks}
\rowcolors{4}{white}{lightgray!40}
\begin{tabular}{lllccc}
\toprule
 \multirow{2}{*}{\rule{0pt}{2ex}\textbf{Attack}} & \multirow{2}{*}{\rule{0pt}{2ex}\textbf{Active/passive}} & \multirow{2}{*}{\rule{0pt}{2ex}\textbf{Property}}& \multicolumn{3}{c}{\textbf{Targets}} \\
                                    \cline{4-6} \rule{0pt}{2ex}       
 \   & \  & \ &  \textbf{Crypto, program flow}   & \textbf{Application inference}  & \textbf{User input} \\
\midrule
Power analysis attacks            & Passive & Physical & \gcmark\ifrefsintab~\cite{DBLP:conf/smartcard/MessergesD99}\fi        & \gcmark\ifrefsintab~\cite{DBLP:conf/internetware/YanGCM15}\fi                         & \rxmark               \\
Electromagnetic analysis attacks  & Passive & Physical &  \gcmark\ifrefsintab~\cite{DBLP:conf/ches/GebotysHT05,DBLP:conf/wistp/NakanoSNSDGKM14,DBLP:conf/cosade/GollerS15,DBLP:conf/ctrsa/BelgarricFMT16,DBLP:conf/ccs/GenkinPPTY16}\fi        & \rxmark                         & \rxmark               \\
Differential computation analysis & Active/passive & Logical &  \gcmark\ifrefsintab~\cite{DBLP:conf/ches/BosHMT16,2015SanfelixWBC}\fi            & \rxmark                         & \rxmark               \\
Smudge attacks                    & Passive & Physical &  \rxmark           & \rxmark                         & \gcmark\ifrefsintab~\cite{DBLP:conf/woot/AvivGMBS10,DBLP:conf/ccs/ZhangXLLLF12,DBLP:conf/wisec/AndriotisTOY13}\fi               \\
Shoulder surfing and reflections  & Passive & Physical &  \rxmark           & \rxmark                         & \gcmark\ifrefsintab~\cite{DBLP:conf/IEEEias/MaggiGB11,DBLP:conf/ccs/RaguramWGMF11,DBLP:journals/tdsc/Raguram0XFGM13,DBLP:conf/ccs/XuH0MF13}\fi               \\
Hand/device movements             & Passive & Physical &  \rxmark           & \rxmark                         & \gcmark\ifrefsintab~\cite{DBLP:conf/ccs/ShuklaKSP14,DBLP:conf/ndss/SunJCZZ016,DBLP:conf/ccs/YueLFLRZ14}\fi               \\
Clock/power glitching             & Active & Physical &  \gcmark\ifrefsintab~\cite{ChipWhispererFIRaspberryPI,DBLP:journals/iacr/OFlynn16}\fi               & \rxmark                         & \rxmark               \\
Electromagnetic fault injection   & Active & Physical &  \gcmark\ifrefsintab~\cite{Ordas2016,DBLP:conf/host/RiviereNRDBS15}\fi         & \rxmark                         & \rxmark               \\
Laser/optical faults              & Active & Physical &  \gcmark\ifrefsintab~\cite{DBLP:conf/ches/SkorobogatovA02,DBLP:conf/fdtc/WoudenbergWM11,DBLP:conf/fdtc/RoscianSDT13}\fi           & \rxmark                         & \rxmark               \\
Temperature variation             & Active & Physical &  \gcmark\ifrefsintab~\cite{DBLP:conf/cardis/HutterS13,DBLP:conf/acns/MullerS13}\fi           & \rxmark                         & \rxmark               \\
NAND mirroring                    & Active & Physical &  \gcmark\ifrefsintab~\cite{DBLP:journals/corr/Skorobogatov16}\fi           & \rxmark                         & \rxmark               \\
\bottomrule
\end{tabular} 
\end{table*}

\attackSection{Laser/Optical Faults}
Optical fault attacks using a laser beam are among the most-effective fault-injection techniques. These attacks take advantage of the fact that a focused laser beam can change the state of a transistor on a microcontroller, resulting in, \eg, bit flips in memory cells. Compared to other fault-injection techniques (voltage glitching, EMFI), the effort for optical fault injection is high. 
First, decapsulation of the chip is a prerequisite in order to access the silicon with the laser beam. Second, finding the correct location for the laser beam to produce exploitable faults is also not a trivial task. 

\attacks
First optical fault-injection attacks targeting an 8-bit microcontroller have been published by Skorobogatov and Anderson \cite{DBLP:conf/ches/SkorobogatovA02} in 2002. Inspired by their work, several optical fault-injection attacks have been published in the following years, most of them targeting smart cards or low-resource embedded devices (\eg, \cite{DBLP:conf/fdtc/WoudenbergWM11}, \cite{DBLP:conf/fdtc/RoscianSDT13}). The increasing number of metal layers on top of the silicon, decreasing feature size (small process technology), and the high decapsulation effort make optical fault injection difficult to apply on modern microprocessors used in smartphones.

\attackSection{Temperature Variation} 
Operating a device outside of its specified temperature range allows to cause faulty behavior. Heating up a device above the maximum specified temperature can cause faults in memory cells. Cooling down the device has an effect on the speed RAM content fades away after power off (\textit{remanence effect} of RAM). 

\attacks
Hutter and Schmidt~\cite{DBLP:conf/cardis/HutterS13} presented heating fault attacks targeting an AVR microcontroller. They prove the practicability of this approach by successfully attacking an RSA implementation on named microcontroller. FROST~\cite{DBLP:conf/acns/MullerS13}, on the other hand, is a tool to recover disc encryption keys from RAM on Android devices by means of cold-boot attacks. Here the authors take advantage of the increased time data in RAM remains valid after power off due to low temperature. 

\attackSection{Differential Computation Analysis}
As already mentioned above, the white-box model assumes that the attacker has full control over the execution environment. This also means that the attacker can produce erroneous or faulty outputs by manipulating intermediate values during the computation. 

\attacks
Sanfelix~\etal~\cite{2015SanfelixWBC} demonstrated that attackers in the white-box model can also perform fault injection attacks. As the attacker has full control over the execution environment and the executed binary, she can also manipulate data during the program execution or manipulate the control flow of the execution. 
Similar to other fault attacks, the idea is to observe differences between normal outputs and erroneous outputs of the binary in order to break the cryptographic implementations. 

\attackSection{NAND Mirroring}
Data mirroring refers to the replication of data storage between different locations. Such techniques are used to recover critical data after disasters but also allow to restore a previous system state. 

\attacks
The Apple iPhone protects a user's privacy by encrypting the data. Therefore, a passcode and a hardware-based key are used to derive various keys that can be used to protect the data on the device. As a dedicated hardware-based key is used to derive these keys, brute-force attempts must be done on the attacked device. Furthermore, brute-force attempts are discouraged by gradually increasing the waiting time between wrongly entered passcodes up to the point where the phone is wiped. In response to the Apple vs FBI case, Skorobogatov~\cite{DBLP:journals/corr/Skorobogatov16} demonstrated that NAND mirroring can be used to reset the phone state and, thus, can be used to brute-force the passcode. Clearly, this approach also represents an active attack as the attacker actively influences (resets) the state of the device. 

\subsection{Overview}
Table~\ref{tab:overview_local_attacks} summarizes the discussed attack categories and the targeted information. 
In terms of targets, we identified cryptographic implementations (crypto), the program flow of applications (which sometimes also allows to attack crypto because different branches might be executed depending on specific key bits), application inference (inference of the executed application), and user input. 
An attack category not targeting specific information (yet), which is indicated by an \rxmark, represents a possible gap that might be investigated in future research. 
For example, power analysis attacks might allow to target user input, such as keystrokes or even actual characters, and shoulder surfing and reflection attacks might well allow to infer running applications.
However, for some attacks it is (highly) unlikely that they will work against specific targets. 
For example, attacking cryptographic algorithms by means of smudge attacks is unlikely to work.

\setlength{\aboverulesep}{0pt}
\setlength{\belowrulesep}{0pt}
\begin{table*}
\centering
\caption{Overview of vicinity side-channel attacks and corresponding targets. \gcmark\ and \rxmark\ indicate whether or not a specific attack has been performed on the corresponding target.}
\label{tab:overview_vicinity_attacks}
\resizebox{\hsize}{!}{
\rowcolors{4}{white}{lightgray!40}
\begin{tabular}{lllcccc}
\toprule
 \multirow{2}{*}{\rule{0pt}{2ex}\textbf{Attack}} & \multirow{2}{*}{\rule{0pt}{2ex}\textbf{Active/passive}} & \multirow{2}{*}{\rule{0pt}{2ex}\textbf{Property}} &  \multicolumn{4}{c}{\textbf{Targets}} \\
                                    \cline{4-7} \rule{0pt}{2ex}       
 \                       & \ & \ & \textbf{Visited websites} & \textbf{Application/action inference}  & \textbf{Identify users/devices} & \textbf{User input} \\
\midrule
Network traffic analysis & Active/passive & Physical/logical & \gcmark\ifrefsintab~\cite{DBLP:conf/cscwd/HeYGLM14,DBLP:conf/ccs/CaiZJJ12,DBLP:conf/wpes/PanchenkoNZE11,DBLP:conf/wpes/WangG13,DBLP:conf/ccs/JuarezAADG14,DBLP:conf/ndss/PanchenkoLPEZHW16}\fi                   & \gcmark\ifrefsintab~\cite{DBLP:journals/tifs/ContiMSV16}\fi                                       & \gcmark\ifrefsintab~\cite{DBLP:conf/wisec/StoberFSM13}\fi                         & \rxmark \\
USB power analysis       & Passive & Physical & \gcmark\ifrefsintab~\cite{Yang2016}\fi                   & \rxmark                                       & \gcmark\ifrefsintab~\cite{DBLP:conf/ccs/ContiNRS16}\fi                         & \rxmark \\
Wi-Fi signal monitoring   & Passive & Physical & \rxmark                   & \rxmark                                       & \rxmark                         & \gcmark\ifrefsintab~\cite{DBLP:journals/mis/ZhangZTXCFLGC16,DBLP:conf/ccs/LiMLZLLR16}\fi \\
\bottomrule
\end{tabular} 
}
\end{table*}

\section{Vicinity Side-Channel Attacks}
 \label{sec:vicinity}
In this section, we survey attacks where the attacker must be in the vicinity of the targeted user/device, \ie, attacks where the attacker compromises, for example, any infrastructure facility within the user's environment.

\subsection{Passive Attacks}
 \label{subsec:vicinity_passive}
\textbf{Network Traffic Analysis.}
In general, the encryption of messages transmitted between two parties only hides the actual content, while specific meta data such as the overall amount of data is not protected. 
This meta data allows to infer sensitive information about the content and the communicating parties. 

\attacks
Network traffic analysis has been extensively studied in the context of website fingerprinting attacks. 
These attacks~\cite{DBLP:conf/ccs/CaiZJJ12,DBLP:conf/wpes/PanchenkoNZE11,DBLP:conf/wpes/WangG13,DBLP:conf/ccs/JuarezAADG14,DBLP:conf/ndss/PanchenkoLPEZHW16} 
wiretap network connections and observe traffic signatures, \eg, unique packet lengths, inter-packet timings, etc., to infer visited websites and even work in case the traffic is routed through Tor. 
While most of these attacks target the network communication in general, attacks explicitly targeting mobile devices also exist. 
For instance, St\"ober~\etal~\cite{DBLP:conf/wisec/StoberFSM13} assumed that an adversary eavesdrops on the UMTS transmission and showed that smartphones can be fingerprinted based on the background traffic of installed apps. 
Conti~\etal~\cite{DBLP:journals/tifs/ContiMSV16} considered an adversary who controls Wi-Fi access points near the targeted device, which allows to infer specific app actions such as posting Facebook status messages. 
In similar settings, traffic analysis techniques allow to fingerprint specific apps as well as actions performed in specific apps~\cite{DBLP:conf/infocom/DaiTWNS13,DBLP:conf/cns/WangYKH15,DBLP:conf/pam/MiskovicLLB15,DBLP:conf/eurosp/TaylorSCM16,DBLP:conf/wisec/AlanK16,DBLP:conf/woot/SaltaformaggioC16}.

While the above presented attacks exploit logical properties, \ie, the fact that encrypted packets do not hide meta data, Schulz~\etal~\cite{DBLP:conf/wisec/SchulzKHT016} exploited the EM emanation of Ethernet cables (hardware properties), which allowed them to observe parts of the transmitted Ethernet frames. 

\attackSection{USB Power Analysis} 
Due to the inherent usage patterns of mobile devices, users are constantly in the need to charge their devices, which is why public USB charging stations have been set up. Similar to power analysis attacks, modified charging stations can be used to collect power traces that allow to infer sensitive information about users and mobile devices. 

\attacks
The identification (or localization) of specific users is considered a privacy risk due to the possibility of tracking individuals. Conti~\etal~\cite{DBLP:conf/ccs/ContiNRS16} demonstrated that wall-socket smart meters that capture the power consumption of plugged devices can be used to identify users/notebooks. Although they demonstrated their attack on notebooks, it is likely that the same attack works for smartphones as well. 
In a similar setting, Yang~\etal~\cite{Yang2016} demonstrated that 
visited websites can be inferred by power traces collected via USB charging stations. 
Such attacks even work if dedicated protection mechanisms, \eg, adapters that block data pins on USB cables, are in place. 

\attackSection{Wi-Fi Signal Monitoring}
Wi-Fi devices continuously monitor the wireless channel (channel state information (CSI)) to effectively transmit data. This is necessary as environmental changes cause the CSI values to change. 

\attacks
Ali~\etal~\cite{DBLP:conf/mobicom/AliLWS15} observed that even finger motions impact wireless signals and cause unique patterns in the time-series of CSI values. In a setting with a sender (notebook) and a receiver (Wi-Fi router), they showed that keystrokes on an external keyboard cause distortions in the Wi-Fi signal. 
They infer entered keys by monitoring these changes of the CSI values. 
Later on, Zhang~\etal~\cite{DBLP:journals/mis/ZhangZTXCFLGC16} inferred unlock patterns on smartphones via a notebook that is connected to the wireless hotspot provided by the smartphone. Li~\etal~\cite{DBLP:conf/ccs/LiMLZLLR16} further improved these attacks by considering an attacker controlling only a Wi-Fi access point. They infer the PIN input on smartphones and also analyze network packets to determine when the sensitive input starts.

\subsection{Active Attacks}
\label{subsec:vicinity_active}
Besides passively observing leaking information, vicinity attacks can be improved by considering active attackers as demonstrated by the following example. 

\attackSection{Network Traffic Analysis} 
Network traffic analysis has already been discussed in the context of passive side-channel attacks. 
Active attackers learn additional information by actively influencing transmitted packets, \eg, by delaying packets. 

\attacks 
He~\etal~\cite{DBLP:conf/cscwd/HeYGLM14} demonstrated that an active attacker, \eg, represented by an Internet Service Provider (ISP), could delay HTTP requests from Tor users in order to increase the performance of website fingerprinting attacks. The idea is that instead of observing the generated traffic for all resources on a webpage in parallel, \ie, the response packets from multiple requests in parallel overlap, an attacker delays the packet requesting a resource until the response from the previous request has been fully retrieved. 

\subsection{Overview}
Table~\ref{tab:overview_vicinity_attacks} summarizes the discussed attack categories and the targeted information. 
The identified targets are the inference of visited websites, application inference (or specific actions within applications), identification of users and devices, and user input.
Again an attack category not targeting specific information (indicated by an \rxmark) represents a possible gap that might be closed in future research. 
For example, USB power analysis attacks might allow to target user input.

\section{Remote Side-Channel Attacks}
 \label{sec:remote}
The attacks presented in this section can be categorized as software-only attacks. In contrast to the local side-channel attacks as well as the vicinity side-channel attacks presented in the previous sections, these attacks neither require the attacker to be in the proximity nor in the vicinity of the targeted user. 
Hence, these attacks can be executed remotely and target a much larger scale since the victim user installed a malicious application on her device. 

\subsection{Passive Attacks}
\label{subsec:remote_passive}
\textbf{Linux-inherited \emph{procfs} Leaks.} 
Linux releases ``accounting'' information that is considered as being harmless via the procfs. This includes, for example, the memory footprint (total virtual memory size and total physical memory size) of each application via \verb|/proc/[pid]/statm|, the CPU utilization times via \verb|/proc/[pid]/stat|, the number of context switches via \verb|/proc/[pid]/status|, but also system-wide information such as interrupt counters via \verb|/proc/interrupts| and context switches via \verb|/proc/stat|. 

\attacks
Jana and Shmatikov~\cite{DBLP:conf/sp/JanaS12a} observed that the memory footprint of the browser correlates with the rendered website. Thus, by monitoring the memory footprint they inferred a user's browsing behavior (browser history), which represents sensitive information and is normally protected by a dedicated permission. 
Later on, Chen~\etal~\cite{DBLP:conf/uss/ChenQM14} exploited this information to detect \emph{Activity} transitions within Android apps. They observed that the shared memory size increases by the size of the graphics buffer in both processes, \ie, the app process and the window compositor process (\emph{SurfaceFlinger}). These increases occur due to the inter-process communication (IPC) between the app and the window manager. Besides, they also considered CPU utilization and network activity in order to infer the exact activity later on. 

Similar to the memory footprint of applications, the procfs also provides system-wide information about the number of interrupts and context switches. Again, this information is considered as being innocuous and is, thus, published on purpose and is accessible without any permission. Simon~\etal~\cite{DBLP:journals/popets/SimonXA16} exploited this information to infer text entered via swipe input methods. More specifically, they observed that the number of interrupts and context switches correlates with the user's finger movements across the keyboard when transitioning from letter to letter. 
Diao~\etal~\cite{DBLP:conf/sp/DiaoLLZ16} presented two attacks to infer unlock patterns and the app running in the foreground. The information leaks exploited were gathered from interrupt time series of the device's touchscreen controller. 
Besides, also the power consumption is released via the procfs. Yan~\etal~\cite{DBLP:conf/internetware/YanGCM15} showed that the power consumption allows to infer the number of entered characters on the soft keyboard.

\attackSection{Data-Usage Statistics}
Android keeps track of the amount of incoming and outgoing network traffic on a per-application basis. These statistics allow users to keep an eye on the data consumption of any app and can be accessed without any permission.

\attacks
Data-usage statistics are captured with a fine-grained granularity, \ie, packet lengths of single TCP packets can be observed, and have already been successfully exploited. Zhou~\etal~\cite{DBLP:conf/ccs/ZhouDHNPWGN13} demonstrated that by monitoring the data-usage statistics an adversary can infer sensitive information of  specific apps. They were able to infer disease conditions accessed via \emph{WebMD}, and the financial portfolio via \emph{Yahoo! Finance}. In addition, they also showed how to infer a user's identity by observing the data-usage statistics of the \emph{Twitter} app and exploiting the publicly available Twitter API.

Later, it has been shown that the data-usage statistics can also be exploited to infer a user's browsing behavior~\cite{DBLP:conf/wisec/SpreitzerGKM16}. The fine-grained statistics of incoming and outgoing network packets allow to fingerprint websites, which even works in case the traffic is routed through the anonymity network Tor.

\attackSection{Page Deduplication} 
To reduce the overall memory footprint of a system, (some) operating systems\footnote{For example, CyanogenMod OS allows to enable page deduplication.} search for identical pages within the physical memory and merge them---even across different processes---which is called page deduplication. As soon as one process intends to write onto such a deduplicated page, a copy-on-write fault occurs and the process gets its own copy of this memory region again. 

\attacks
Such copy-on-write faults have been exploited by Suzaki~\etal~\cite{DBLP:conf/eurosec/SuzakiIYA11} to detected applications on Linux and Windows as well as file downloads. 
Recently, Gruss~\etal~\cite{DBLP:conf/esorics/GrussBM15} demonstrated the possibility to measure the timing differences between normal write accesses and copy-on-write faults from within JavaScript code. Based on these precise timings they suggest to fingerprint visited websites by allocating memory that stores images found on popular websites. If the user browses the website with the corresponding image, then at some point the OS detects the identical content in the pages and deduplicates these pages. By continuously writing to the allocated memory, the attacker might observe a copy-on-write fault in which case the attacker knows that the user currently browses the corresponding website. 

\attackSection{Microarchitectural Attacks}
Modern computer architectures include many components to improve the overall effectiveness and performance. For instance, CPU caches represent an important component within the memory hierarchy of modern computer architectures. Multiple cache levels bridge the gap between the latency of main memory accesses and the fast CPU clock frequencies. 
Microarchitectural attacks exploit specific effects like the timing behavior of these components, \eg, branch prediction units and CPU caches, in order to learn sensitive information about executed instructions, code paths, etc. More specifically, by measuring execution times and memory accesses, an attacker can infer sensitive information from processes running in parallel on the same device. As CPU caches have been shown to represent a powerful source of information leaks, we focus on cache attacks. 

\attacks
Cache-timing attacks against  AES have already been investigated on Android-based mobile devices. For instance, Bernstein's cache-timing attack~\cite{2004-bernstein-cachetiming} has been launched on development boards \cite{DBLP:conf/fc/WeissHS12,DBLP:conf/intrust/WeissWAS14,DBLP:conf/esorics/ZanklMHS16} and on Android smartphones~\cite{DBLP:conf/nss/SpreitzerP13,DBLP:conf/wistp/SpreitzerG14} in order to reduce the effective key size of AES. Besides, similar cache attacks have been launched on embedded devices~\cite{DBLP:conf/ctrsa/BogdanovEPW10} and more fine-grained attacks~\cite{DBLP:journals/joc/TromerOS10} against AES have also been applied on smartphones~\cite{DBLP:conf/cosade/SpreitzerP13}. These attacks relied on privileged access to precise timing measurements, but as stated by Oren~\etal~\cite{DBLP:conf/ccs/OrenKSK15} cache attacks can also be exploited via JavaScript and, thus, do not require native code execution anymore. They even demonstrated the possibility to track user behavior including mouse movements as well as browsed websites via JavaScript-based cache attacks. A recent paper by Lipp~\etal~\cite{DBLP:conf/uss/LippGSMM16} demonstrates that all existing cache attacks, including the effective Flush+Reload attack~\cite{DBLP:conf/uss/YaromF14}, can be applied on modern Android smartphones without any privileges. While early attacks on smartphones exclusively targeted cryptographic implementations, their work also shows that user interactions (touch actions and swipe actions) can be inferred through this side channel. Similar investigations of Flush+Reload on ARM have also been conducted by Zhang~\etal~\cite{DBLP:conf/ccs/ZhangXZ16}.

As some of these attacks actively influence the behavior of the victim, \eg, the execution time, some microarchitectural attacks can also be considered as active attacks. 
For a more detailed survey about microarchitectural attacks in general, we refer to the survey papers by Ge~\etal~\cite{Ge2016} and Szefer~\cite{DBLP:journals/iacr/Szefer16}.

\attackSection{Sensor-based Keyloggers}
Cai~\etal~\cite{DBLP:conf/sigcomm/CaiMC09} and Raij~\etal~\cite{DBLP:conf/chi/RaijGKS11} were one of the first to discuss privacy implications resulting from mobile devices equipped with cameras, microphones, GPS sensors, and motion sensors in general. Nevertheless, a category of attacks that received significant attention are sensor-based keyloggers. These attacks are based on two observations. First, smartphones are equipped with lots of sensors---both motion sensors as well as ambient sensors---that can be accessed without any permission, and second, these devices are operated with fingers while being held in the users' hands. Hence, the following attacks are all based on the observation that users tap/touch/swipe the touchscreen and that the device is slightly tilt and turned during the operation.

\attacks 
In 2011, Cai and Chen~\cite{DBLP:conf/uss/CaiC11} were the first to observe a correlation between entered digits on touchscreens and the readings from the accelerometer sensor that can be exploited for motion-based keylogging attacks. 
Following this work, Owusu~\etal~\cite{DBLP:conf/wmcsa/OwusuHDPZ12} extended the attack to infer single characters, and Aviv~\cite{Aviv2012SideChannelsSmartphones} and Aviv~\etal~\cite{DBLP:conf/acsac/AvivSBS12} investigated the accelerometer to attack PIN and pattern inputs. Subsequent publications~\cite{DBLP:conf/wisec/XuBZ12,DBLP:conf/trust/CaiC12,DBLP:conf/mobisys/MiluzzoVBC12} also considered the combination of the accelerometer and the gyroscope in order to improve the performance as well as to infer even longer text inputs~\cite{DBLP:conf/wisec/PingSM15}.

Since the W3C specifications allow access to the motion and orientation sensors from JavaScript, motion-based keylogging attacks have even been performed via websites~\cite{DBLP:journals/istr/MehrnezhadTSH16,DBLP:journals/corr/MehrnezhadTSH16a}. Even worse, some browsers continue to execute JavaScript, although the user closed the browser or turned off the screen. 

While the above summarized attacks exploit different motion sensors, \eg, accelerometer and gyroscope, ambient sensors can also be used for keylogging attacks. 
Spreitzer~\cite{DBLP:conf/ccs/Spreitzer14} presented an attack that exploits an ambient sensor, namely the ambient-light sensor, in order to infer a user's PIN input on touchscreens. Minor tilts and turns during keyboard input lead to variations of the ambient-light sensor readings, which are then correlated with keyboard input on the touchscreen.

As demonstrated by Simon and Anderson~\cite{DBLP:conf/ccs/SimonA13}, PIN inputs on smartphones can also be inferred by continuously taking pictures via the front camera. Afterwards, PIN digits can be inferred by image analysis and by investigating the relative changes of objects in subsequent pictures that correlate with the entered digits. Fiebig~\etal~\cite{DBLP:conf/woot/FiebigKH14} demonstrated that the front camera can be used to capture the screen reflections in the user's eyeballs, which allows to infer user input. In a similar manner, Narain~\etal~\cite{DBLP:conf/wisec/NarainSN14} and Gupta~\etal~\cite{DBLP:conf/dbsec/GuptaSAV16} showed that tap sounds (inaudible to the human ear) recorded via smartphone stereo-microphones can be used to infer typed text on the touchscreen. However, these attacks require 
dedicated permissions to access the camera and the microphone, which might raise the user's suspicion. 
In contrast, the motion and ambient sensors can be accessed without any permission. 

For a more complete overview of sensor-based keylogging attacks, we refer to the survey papers by Hussain~\etal~\cite{DBLP:journals/percom/HussainAZZKAA16} and Nahapetian~\cite{DBLP:conf/ccnc/Nahapetian16}. Considering the significant number of papers that have been published in this context, user awareness about such attacks should be raised. Especially since Mehrnezhad~\etal~\cite{DBLP:journals/corr/MehrnezhadTSH16a} found that the perceived risk of motion sensors, especially ambient sensors, among users is very low.

\attackSection{Fingerprinting Devices/Users}
The identification of smartphones (and users) without a user's awareness is considered a privacy risk. While obvious identification mechanisms such as device IDs and web cookies can be thwarted, imperfections of hardware components, \eg, sensors, as well as specific software features can also be employed to stealthily fingerprint and identify devices and users, respectively. 

\attacks 
Bojinov~\etal~\cite{DBLP:journals/corr/BojinovMNB14} and Dey~\etal~\cite{DBLP:conf/ndss/DeyRXCN14} observed that unique variations of sensor readings (\eg, of the accelerometer) can be used to fingerprint devices. These variations are a result of the manufacturing process and are persistent throughout the life of the sensor/device. As these sensors can be accessed via JavaScript, it is possible to fingerprint devices via websites~\cite{DBLP:conf/ndss/DasBC16}. Similarly, such imperfections also affect the microphones and speakers~\cite{DBLP:conf/ccs/DasBC14,DBLP:conf/ccs/ZhouDLZ14}, which also allow to fingerprint devices. In addition, by combining multiple sensors, even higher accuracies can be achieved~\cite{DBLP:conf/dimva/HupperichHH16}. 

Kurtz~\etal~\cite{DBLP:journals/popets/KurtzGBRF16} demonstrated how to fingerprint mobile device configurations, \eg, device names, language settings, installed apps, etc. Hence, their fingerprinting approach exploits software properties (\ie, software configurations) only. Hupperich~\etal~\cite{DBLP:conf/acsac/HupperichMKHG15} proposed to combine hardware features as well as software features to fingerprint mobile devices.

\newcommand*\rot{\rotatebox{0}}
\setlength{\aboverulesep}{0pt}
\setlength{\belowrulesep}{0pt}
\begin{table*}
\centering
\caption{Overview of remote side-channel attacks and corresponding targets. \gcmark\ and \rxmark\ indicate whether or not a specific attack has been performed on the corresponding target.}
\label{tab:overview_remote_attacks}
\resizebox{\hsize}{!}{
\rowcolors{5}{lightgray!40}{white}
\begin{tabular}{lllccccccc}
\toprule
 \multirow{3}{*}{\rule{0pt}{2ex}\textbf{Attack}} & \multirow{3}{*}{\rule{0pt}{2ex}\textbf{Active/passive}} & \multirow{3}{*}{\rule{0pt}{2ex}\textbf{Property}} &\multicolumn{7}{c}{\textbf{Targets}} \\
                                    \cline{4-10} \rule{0pt}{2ex}       
 \                           & \ & \ & \textbf{Visited} & \textbf{Application/action}  & \textbf{Identify} & \multirow{2}{*}{\rule{0pt}{2ex}\textbf{User input}} & \multirow{2}{*}{\rule{0pt}{2ex}\textbf{Crypto}}  & \rot{\textbf{Location}} & \rot{\textbf{Privilege}}\\
 \ & \ & \ & \textbf{websites} & \textbf{inference} & \textbf{users/devices} & \ & \ & \textbf{inference} & \textbf{escalation}\\
\midrule
procfs leaks                 & Passive & Physical/logical & \gcmark\ifrefsintab~\cite{DBLP:conf/sp/JanaS12a,DBLP:conf/wisec/SpreitzerGKM16}\fi                   & \gcmark\ifrefsintab~\cite{DBLP:conf/uss/ChenQM14,DBLP:conf/sp/DiaoLLZ16}\fi                                       & \gcmark\ifrefsintab~\cite{DBLP:conf/ccs/ZhouDHNPWGN13}\fi                         & \gcmark\ifrefsintab~\cite{DBLP:journals/popets/SimonXA16,DBLP:conf/sp/DiaoLLZ16,DBLP:conf/internetware/YanGCM15}\fi             & \rxmark          & \gcmark\ifrefsintab~\cite{DBLP:conf/uss/MichalevskySVBN15}\fi & \rxmark \\
Data-usage statistics        & Passive & Logical &  \gcmark\ifrefsintab~\cite{DBLP:conf/wisec/SpreitzerGKM16}\fi                   & \gcmark\ifrefsintab~\cite{DBLP:conf/ccs/ZhouDHNPWGN13}\fi                                       & \gcmark\ifrefsintab~\cite{DBLP:conf/ccs/ZhouDHNPWGN13}\fi                         & \rxmark             & \rxmark          & \rxmark & \rxmark \\
Page deduplication           & Passive & Logical &  \gcmark\ifrefsintab~\cite{DBLP:conf/esorics/GrussBM15}\fi                   & \gcmark\ifrefsintab~\cite{DBLP:conf/eurosec/SuzakiIYA11}\fi                                       & \rxmark                         & \rxmark             & \rxmark          & \rxmark & \rxmark \\
Microarchitectural attacks   & Active/passive & Physical &  \gcmark\ifrefsintab~\cite{DBLP:conf/ccs/OrenKSK15}\fi                   & \gcmark\ifrefsintab~\cite{DBLP:conf/ccs/OrenKSK15,DBLP:conf/uss/LippGSMM16}\fi                                       & \rxmark                         & \gcmark\ifrefsintab~\cite{DBLP:conf/uss/LippGSMM16}\fi             & \gcmark\ifrefsintab~\cite{DBLP:conf/uss/LippGSMM16,DBLP:conf/nss/SpreitzerP13,DBLP:conf/wistp/SpreitzerG14,DBLP:conf/ctrsa/BogdanovEPW10,DBLP:conf/cosade/SpreitzerP13}\fi          & \rxmark & \rxmark \\
Sensors                      & Passive & Physical &  \rxmark                   & \gcmark                                       & \gcmark\ifrefsintab~\cite{DBLP:journals/corr/BojinovMNB14,DBLP:conf/ndss/DeyRXCN14,DBLP:conf/ndss/DasBC16}                         & \gcmark\ifrefsintab~\cite{DBLP:conf/uss/CaiC11,DBLP:conf/wmcsa/OwusuHDPZ12,Aviv2012SideChannelsSmartphones,DBLP:conf/acsac/AvivSBS12,DBLP:conf/wisec/XuBZ12,DBLP:conf/trust/CaiC12,DBLP:conf/mobisys/MiluzzoVBC12,DBLP:conf/wisec/PingSM15,DBLP:journals/istr/MehrnezhadTSH16,DBLP:journals/corr/MehrnezhadTSH16a,DBLP:conf/ccs/Spreitzer14,DBLP:conf/uss/MichalevskyBN14}             & \rxmark          & \gcmark\ifrefsintab~\cite{DBLP:conf/sensys/Ho0SS15}\fi & \rxmark \\
Microphone                   & Passive & Physical &  \rxmark                   & \rxmark                                       & \gcmark\ifrefsintab~\cite{DBLP:conf/ccs/DasBC14,DBLP:conf/ccs/ZhouDLZ14}\fi                         & \gcmark \ifrefsintab~\cite{DBLP:conf/wisec/NarainSN14,DBLP:conf/dbsec/GuptaSAV16,DBLP:conf/ndss/SchlegelZZIKW11}\fi             & \rxmark          & \rxmark & \rxmark \\
Speakers                     & Passive & Physical &  \rxmark                   & \rxmark                                       & \gcmark\ifrefsintab~\cite{DBLP:conf/ccs/DasBC14,DBLP:conf/ccs/ZhouDLZ14}\fi                         & \rxmark             & \rxmark          & \rxmark & \rxmark \\
Camera                       & Passive & Physical &  \rxmark                   & \rxmark                                       & \rxmark                         & \gcmark \ifrefsintab~\cite{DBLP:conf/ccs/SimonA13,DBLP:conf/woot/FiebigKH14}\fi             & \rxmark          & \rxmark & \rxmark \\
Device configurations        & Passive & Logical &  \rxmark                   & \rxmark                                       & \gcmark\ifrefsintab~\cite{DBLP:journals/popets/KurtzGBRF16}\fi                         & \rxmark             & \rxmark          & \rxmark & \rxmark \\
Rowhammer         & Active & Physical &  \rxmark                   & \rxmark                                       & \rxmark & \rxmark             & \rxmark          & \rxmark & \gcmark \ifrefsintab~\cite{DBLP:conf/ccs/VeenFLGMVBRG16}\fi\\
\bottomrule
\end{tabular} 
}
\end{table*}

\attackSection{Location Inference}
As smartphones are always carried around, information about a phone's location inevitably reveals the user's location. Hence, resources that obviously can be used to determine a user's location, \eg, the GPS sensor, are considered as privacy relevant and, thus, require a dedicated permission. 
Yet, even without permissions, side-channel attacks can be used to infer precise location information about users. 

\attacks 
Han~\etal~\cite{DBLP:conf/comsnets/HanONPZ12}, Nawaz~\etal~\cite{DBLP:conf/sensys/NawazM14}, and Narain~\etal~\cite{DBLP:conf/sp/NarainVBN16} demonstrated that the accelerometer and the gyroscope can be used to infer car driving routes. Similarly, Hemminki~\etal~\cite{DBLP:conf/sensys/HemminkiNT13} 
inferred the transportation mode, \eg, train, bus, metro, etc., via the accelerometer readings. 
Besides the accelerometer and the gyroscope, ambient sensors can also be used to infer driving routes. Ho~\etal~\cite{DBLP:conf/sensys/Ho0SS15} exploited the correlation between sensor readings of the barometer sensor and the geographic elevation to infer driving routes. 

Even less obvious side-channels that allow to infer driving routes and locations are the speaker status information (\eg, speaker on/off) and the power consumption (available via the procfs). More specifically, Zhou~\etal~\cite{DBLP:conf/ccs/ZhouDHNPWGN13} observed that the Android API allows to query whether or not the speaker is currently active, \ie, boolean information that indicates whether or not any app is playing sound on the speakers. They exploit this information to attack the turn-by-turn voice guidance of navigation systems. By continuously querying this API, they determine how long the speaker is active. This information allows to infer the speech length of voice direction elements, \eg, the length of ``Turn right onto East Main Street''. As driving routes consist of many such turn-by-turn voice guidances, fingerprinting driving routes is possible. 

Michalevsky~\etal~\cite{DBLP:conf/uss/MichalevskySVBN15} observed that the power consumption (available in the procfs) is related to the strength of the cellular signal, which depends on the distance to the base station. Given this information, they inferred a user's location. 

\attackSection{Speech Recognition} 
Eavesdropping conversations represents a severe privacy threat. Thus, a dedicated permission protects the access to the microphone. 
However, acoustic signals, such as human speech, in the vicinity of a mobile device also influence the gyroscope measurements. 

\attacks
Michalevsky~\etal~\cite{DBLP:conf/uss/MichalevskyBN14} exploited the gyroscope sensor to measure acoustic signals in the vicinity of the phone and to recover speech information. Although they only consider a small set of vocabulary, \ie, digits only, their work demonstrates the immense power of gyroscope sensors in today's smartphones. By exploiting the gyroscope sensor to eavesdrop on a user's conversations they are able to bypass the permission required to access the microphone.

\attackSection{Soundcomber} 
Customer service departments often rely on automated menu services to interact with customers over the phone. A well-known example are the interactive voice response systems supported by telephone services that use dual-tone multi-frequency (DTMF) signaling to transmit entered numbers, \ie, an audio signal is transmitted for each key. 

\attacks
As DTMF tones are also played locally, Schlegel~\etal~\cite{DBLP:conf/ndss/SchlegelZZIKW11} showed that by requesting permission to access the microphone, 
these tones can be recorded and used to infer sensitive input provided to these automated menu services. More specifically, they exploit this information to infer credit card numbers entered while interacting with such interactive voice response systems of credit card companies.

\subsection{Active Attacks}
\label{subsec:remote_active}
An area of research that gains increasing attention among the scientific community are active side-channel attacks that can be exploited via software execution only. The most prominent example is the so-called Rowhammer attack that exploits DRAM disturbance errors to conduct software-induced fault attacks. 

\attackSection{Rowhammer} 
The increasing density of memory cells within the DRAM requires the size of these cells to decrease, which in turn decreases the charging of single cells but also causes electromagnetic coupling effects between cells. 

\attacks
Kim~\etal~\cite{DBLP:conf/isca/KimDKFLLWLM14} demonstrated that these observations can be used to induce hardware faults, \ie, bit flips in neighboring cells, via frequent memory accesses to the main memory. Thereby, they showed that frequent memory accesses in the attacker's memory allow to induce faults (bit flips) in the victim's memory. 
Seaborn and Dullien~\cite{2015SeabornRowhammer} demonstrated how to possibly exploit these bit flips from native code and Gruss~\etal~\cite{DBLP:conf/dimva/GrussMM16} showed that such bit flips can even be induced via JavaScript code. A recent paper~\cite{DBLP:conf/ccs/VeenFLGMVBRG16} demonstrates the exploitation of the Rowhammer bug to gain root privileges on Android smartphones by inducing bit flips from an unprivileged application.

\subsection{Overview}
Table~\ref{tab:overview_remote_attacks} summarizes the discussed attack categories and the targeted information. 
The target ``application/action inference'' also refers to sensitive information that can be inferred from specific actions. 
For example, diseases conditions, stock portfolios, etc. can be inferred from data-usage statistics (cf.~\cite{DBLP:conf/ccs/ZhouDHNPWGN13}).
The target ``user input'' refers to PIN and pattern inputs on the screen, inter-keystroke timing information, and also the DTMF tone exploitation~\cite{DBLP:conf/ndss/SchlegelZZIKW11}.
Again an attack category not targeting specific information (yet), which is indicated by an \rxmark, represents a possible gap that might be closed in future research.

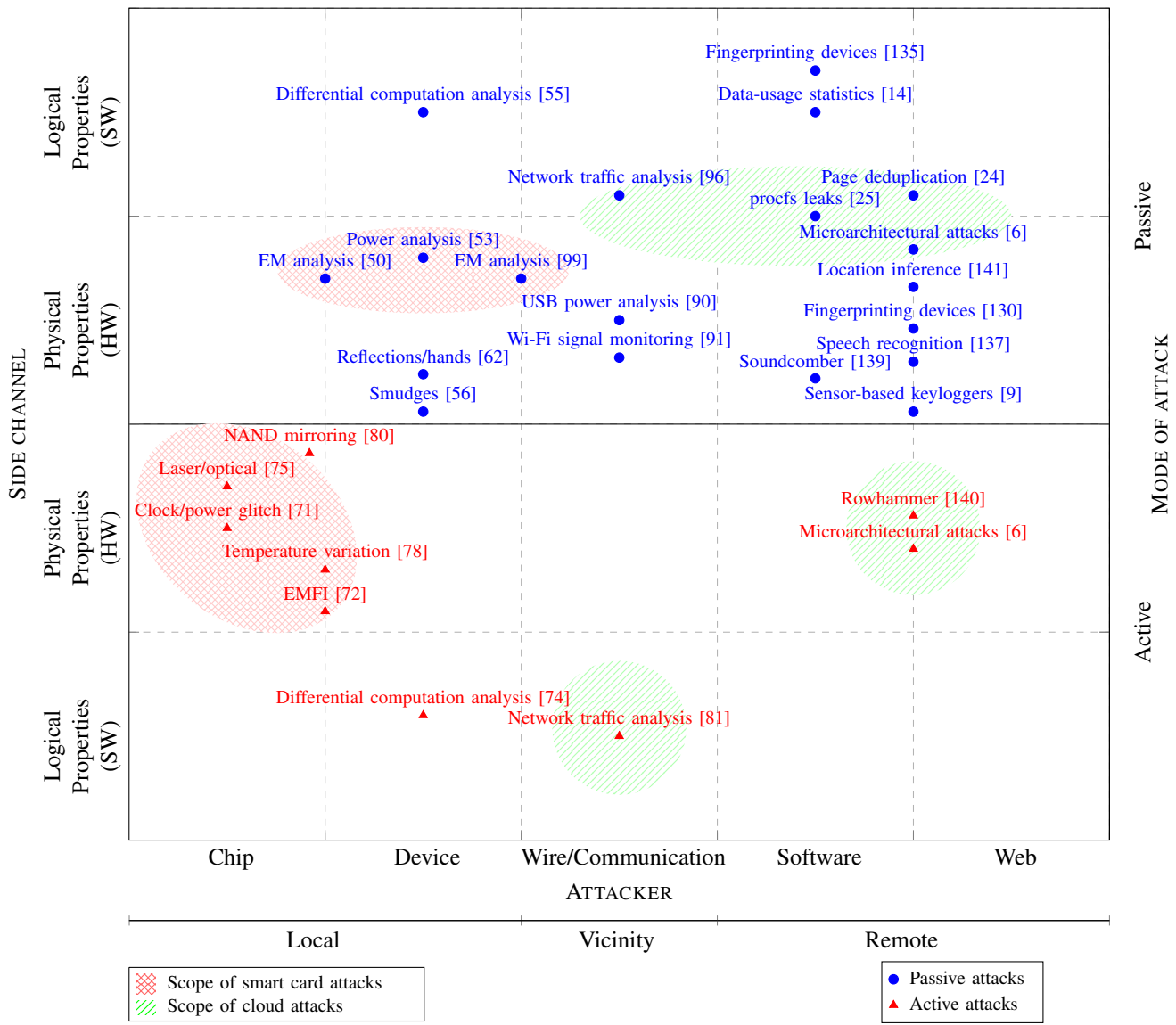
\begin{figure*}
 \centering
 \pgfplotsset{ every non boxed x axis/.append style={x axis line style=-} }
  \begin{tikzpicture}
    \begin{axis}[ylabel={\textsc{Side channel}},xlabel={\textsc{Attacker}},
                y label style={at={(axis description cs:-0.03,.5)},rotate=0},
	        width=16.2cm,height=14cm,
	        disabledatascaling,
		grid=major,
		axis lines =box, 
		grid style={dashed,gray!60},
		ymin=-2,ymax=2,
		extra y ticks=0,
                extra y tick style={grid=major, grid style={solid,black}},
		xmin=0,xmax=5,
	        xtick={1,2,3,4,5},
		xticklabels={Chip,Device,Wire/Communication,Software,Web},
		xticklabel style = {xshift=-1.4cm},
		ytick={1,2,0.001,-1},
		yticklabels={Physical\\ Properties\\(HW), Logical \\Properties\\(SW),, Logical \\Properties\\(SW)}, 
		yticklabel style = {yshift=-1.7cm,rotate=90, align=center},
		]
     \end{axis}
     \begin{axis}[yshift=-1.2cm,axis y line=none, axis x line=bottom, 
		  anchor=origin,
		  width=16.2cm,height=14cm,
		  axis lines =box,
		  axis line style={},
		  axis x line=middle,axis y line=middle,
		  ymin=0,ymax=2,
		  xmin=0,xmax=5,
		  xtick={2,5},
		  xticklabels={Local,Remote},
		  xticklabel style={yshift=1,xshift=-3.1cm,align=center},
     ]      
     \end{axis}
     \begin{axis}[yshift=-1.2cm,axis y line=none, axis x line=bottom, 
		  anchor=origin,
		  width=16.2cm,height=14cm,
		  axis lines =box,
		  axis line style={},
		  axis x line=middle,axis y line=middle,
		  ymin=0,ymax=2,
		  xmin=0,xmax=5,
		  xtick={0.001,3},
		  xticklabels={,Vicinity},
		  xticklabel style={yshift=1,xshift=-1.5cm,align=center},
     ]      
     \end{axis}
     \begin{axis}[ylabel = {\textsc{Mode of attack}},
        ylabel style = {yshift=-16.6cm,align=center},
        axis y line*=right,
	axis x line=none,
	width=16.2cm,height=14cm,
	ymin=0,ymax=2,
	anchor=origin,
	xmin=0,xmax=5,
        ytick={.5,1.5},
        yticklabels={Active, Passive},
	yticklabel style = {xshift=.25cm,rotate=90, align=center},
	legend style={at={(.85,-.22)},anchor=south}
        ]
     \node (SPA)    at (125,135) {};
     \node (EM)     at ( 65,135) {};
     \node (Laser)  at ( 50, 70) {};
     
     \node (NW)     at (230,160) {};
     \node (NW1)    at (230,150) {};
     \node (Dedup)  at (450,160) {};
     \node (Cache1) at (495,110) {};
     \node (Cache2) at (465,110) {};
     \node (Row)    at (380, 55) {};
     \node (Row1)   at (380, 75) {};
\addplot[blue,mark=*,mark options={fill=blue},nodes near coords,only marks,
   point meta=explicit symbolic,
   visualization depends on={value \thisrow{anchor}\as\myanchor},
   every node near coord/.append style={anchor=\myanchor}
] 
table[meta=label] {
x y label anchor

1.5 1.75 {\footnotesize Differential computation analysis\ifrefsinfig~\cite{DBLP:conf/ches/BosHMT16}\fi} south

1.5 1.03 {\footnotesize Smudges\ifrefsinfig~\cite{DBLP:conf/woot/AvivGMBS10}\fi} south
1.5 1.12 {\footnotesize Reflections/hands\ifrefsinfig~\cite{DBLP:conf/ccs/RaguramWGMF11}\fi} south

 1 1.35 {\footnotesize EM analysis\ifrefsinfig~\cite{DBLP:conf/ches/GebotysHT05}\fi} south
1.5 1.4 {\footnotesize Power analysis\ifrefsinfig~\cite{DBLP:conf/ctrsa/BelgarricFMT16}\fi} south

2 1.35 {\footnotesize EM analysis\ifrefsinfig~\cite{DBLP:conf/wisec/SchulzKHT016}\fi} south

4 1.03 {\footnotesize Sensor-based keyloggers\ifrefsinfig~\cite{DBLP:conf/uss/CaiC11}\fi} south
4 1.15 {\footnotesize Speech recognition\ifrefsinfig~\cite{DBLP:conf/uss/MichalevskyBN14}\fi} south
4 1.23 {\footnotesize Fingerprinting devices\ifrefsinfig~\cite{DBLP:conf/ndss/DeyRXCN14}\fi} south

3.5 1.11 {\footnotesize Soundcomber\ifrefsinfig~\cite{DBLP:conf/ndss/SchlegelZZIKW11}\fi} south
4 1.33 {\footnotesize Location inference\ifrefsinfig~\cite{DBLP:conf/comsnets/HanONPZ12}\fi} south
4 1.42 {\footnotesize Microarchitectural attacks\ifrefsinfig~\cite{DBLP:conf/uss/YaromF14}\fi} south
3.5 1.5 {\footnotesize procfs leaks\ifrefsinfig~\cite{DBLP:conf/sp/JanaS12a}\fi} south

4 1.55 {\footnotesize Page deduplication\ifrefsinfig~\cite{DBLP:conf/esorics/GrussBM15}\fi} south

3.5 1.85 {\footnotesize Fingerprinting devices\ifrefsinfig~\cite{DBLP:journals/popets/KurtzGBRF16}\fi} south
3.5 1.75 {\footnotesize Data-usage statistics\ifrefsinfig~\cite{DBLP:conf/ccs/ZhouDHNPWGN13}\fi} south
2.5 1.55 {\footnotesize Network traffic analysis\ifrefsinfig~\cite{DBLP:conf/eurosp/TaylorSCM16}\fi} south
2.5 1.25  {\footnotesize USB power analysis\ifrefsinfig~\cite{DBLP:conf/ccs/ContiNRS16}\fi} south
2.5 1.16  {\footnotesize Wi-Fi signal monitoring\ifrefsinfig~\cite{DBLP:journals/mis/ZhangZTXCFLGC16}\fi} south
};

\addplot[red, mark=triangle*,mark options={solid,fill=red},nodes near coords,only marks,
   point meta=explicit symbolic,
   visualization depends on={value \thisrow{anchor}\as\myanchor},
   every node near coord/.append style={anchor=\myanchor}
] 
table[meta=label] {
x y label anchor
.5 .85 {\footnotesize Laser/optical\ifrefsinfig~\cite{DBLP:conf/ches/SkorobogatovA02}\fi} south
.5 .75 {\footnotesize Clock/power glitch\ifrefsinfig~\cite{DBLP:journals/iacr/OFlynn16}\fi} south
1 .65 {\footnotesize Temperature variation\ifrefsinfig~\cite{DBLP:conf/cardis/HutterS13}\fi} south
1 .55 {\footnotesize EMFI\ifrefsinfig~\cite{Ordas2016}\fi} south
.92 .93 {\footnotesize NAND mirroring\ifrefsinfig~\cite{DBLP:journals/corr/Skorobogatov16}\fi} south

1.5 .3 {\footnotesize Differential computation analysis\ifrefsinfig~\cite{2015SanfelixWBC}\fi} south

2.5 .25 {\footnotesize Network traffic analysis\ifrefsinfig~\cite{DBLP:conf/cscwd/HeYGLM14}\fi} south

4 0.78 {\footnotesize Rowhammer\ifrefsinfig~\cite{DBLP:conf/ccs/VeenFLGMVBRG16}\fi} south
4 0.7 {\footnotesize Microarchitectural attacks\ifrefsinfig~\cite{DBLP:conf/uss/YaromF14}\fi} south
};
\addlegendentry{\footnotesize ~Passive attacks~~}
\addlegendentry{\footnotesize ~Active attacks~~~}

%
  \def\firstSmart{(150,137) ellipse [x radius=22, y radius=35, rotate=90]}
  \def\secondSmart{(60,75) ellipse [x radius=62, y radius=22, rotate=140]}
  \def\firstCloud{(340,150) ellipse [x radius=110, y radius=12, rotate=180]}
  \def\secondCloud{(400,75) circle (1cm)}
  \def\thirdCloud{(250,27) circle (1cm)}
  \fill[white,postaction={pattern=crosshatch, pattern color=red!20!white}] \firstSmart \secondSmart; 
  \fill[white,postaction={pattern=north east lines, pattern color=green!30!white}] \firstCloud \secondCloud \thirdCloud;
   
     \end{axis} 
  \fill[white,postaction={pattern=crosshatch, pattern color=red!60!white}]   (.1,-2) rectangle (.4,-2.3) node[pos=0.5,right,black] (R1) {\footnotesize{~~Scope of smart card attacks}};
  \fill[white,postaction={pattern=north east lines, pattern color=green!100!white}] (.1,-2.4) rectangle (.4,-2.6) node[pos=0.5,right,black]{\footnotesize{~~Scope of cloud attacks}};

    \draw (0,-1.9) -| (4.4,-2.7) -| (0,-1.9);
 \end{tikzpicture}
\caption{Overview of side-channel attacks: (1) active vs passive, (2) logical properties vs physical properties, (3) local vs vicinity vs remote.}
  \label{fig:big_picture}
\end{figure*}

\section{Trend Analysis}
 \label{sec:big_picture}
In Figure~\ref{fig:big_picture} we classify the attacks surveyed in Sections~\ref{sec:local}--\ref{sec:remote} according to our new classification system. 
We distinguish between \emph{active} and \emph{passive} attackers along the (right) y-axis. Passive attacks are classified above the x-axis and active attacks are classified below the x-axis. The (left) y-axis distinguishes between the exploitation of \emph{physical properties} and \emph{logical properties}.
As both of these categories can be exploited by passive as well as active attackers, these two categories are mirrored along the x-axis. The x-axis categorizes side-channel attacks according to the attacker's proximity to the targeted device. For instance, some attacks require an attacker to have access to the targeted device or even to have access to components within the device, \eg, the attacker might remove the back cover in order to measure the EM emanation of the chip. Stronger adversaries (with weaker assumptions) might rely on wiretapping techniques. The strongest adversaries rely on unprivileged applications being executed on the targeted device or even only that the victim visits a malicious website. 

Based on this classification system we observe specific trends in modern side-channel attacks that will be discussed within the following paragraphs. This trend analysis also includes pointers for possible research directions. 

\attackSection{From Local to Remote Attacks}
The first trend that can be observed is that, in contrast to the smart card era, the smartphone era faces a shift towards remote side-channel attacks that focus on both hardware properties and software features. The shift from local attacks (during the smart card era) towards remote attacks (on mobile devices) can be addressed to the fact that the attack scenario as well as the attacker have changed significantly. More specifically, side-channel attacks against smart cards have been conducted to reveal sensitive information that should be protected from being accessed by benign users. For example, in case of pay-TV cards the secret keys must be protected against benign users, \ie, users who bought these pay-TV cards in the first place. The attacker in this case might be willing to invest in equipment in order to reveal the secret key as this key could be sold later on. 

In contrast, today's smartphones are used to store and process sensitive information, and attackers interested in this information are usually not the users themselves but rather criminals, imposters, and other malicious entities that aim to steal this sensitive information from users. Especially the appification of the mobile ecosystem provides tremendous opportunities for attackers to exploit identified side-channel leaks via software-only attacks. Hence, this shift also significantly increases the scale at which attacks are conducted. While local attacks only target a few devices, remote attacks can be conducted on millions of devices at the same time by distributing software via available app markets.

\attackSection{From Active to Passive Attacks}
The second trend that can be observed is that fault injection attacks have been quite popular on smart cards, whereas such (local) fault attacks are not that widely investigated on smartphones, at least at the moment. Consequently, we also observe that the variety of fault attacks conducted in the smart card era has decreased significantly in the smartphone era, which can be addressed to the following observations. First, the targeted device itself, \eg, a smartphone, is far more expensive than a smart card and, hence, fault attacks that potentially permanently break the device are only acceptable for very targeted attacks. Even in case of highly targeted attacks (cf. Apple vs FBI dispute), zero-day vulnerabilities might be chosen instead of local fault attacks.\footnote{However, in September 2016 Skorobogatov~\cite{DBLP:journals/corr/Skorobogatov16} demonstrated that NAND mirroring allows to bypass the PIN entry limit on the iPhone 5c.} Second, remote fault attacks seem to be harder to conduct as such faults are harder to induce via software execution. Currently, the only remote fault attack (also known as software-induced fault attack) is the Rowhammer attack, which however gets increasing attention among the scientific community and has already been exploited to gain root access on Android devices~\cite{DBLP:conf/ccs/VeenFLGMVBRG16}. Although software-induced fault attacks have not been investigated extensively in the past, we expect further research to be conducted in this context. 

Some microarchitectural attacks can also be considered as active attacks because the attacker influences the behavior of the targeted program (victim). 
For example, cache attacks can be used to slow down the execution of the victim due to cache contention. However, this does not introduce a fault in the computation and, hence, Rowhammer currently represents the only software-induced fault attack.

\attackSection{Exploiting Physical and Logical Properties}
In contrast to physical properties, logical properties (software features) do not result from any physical interaction with the device, but due to dedicated features provided via software. While traditional side-channel attacks mostly exploited physical properties and required dedicated equipment, more recent side-channel attacks exploit physical properties as well as logical properties. Interestingly, the immense number of sensors in smartphones also allows to exploit physical properties by means of software-only attacks, which was not possible on smart cards. Although the majority of attacks on mobile devices still exploits physical properties, the exploitation of logical properties also receives increasing attention. Especially the procfs seems to provide an almost inexhaustible source for possible information leaks. For example, the memory footprint released via the procfs has been used to infer visited websites~\cite{DBLP:conf/sp/JanaS12a}, or the number of context switches has been used to infer swipe input~\cite{DBLP:journals/popets/SimonXA16}. Besides, information that is available via official APIs is in some cases also available via the procfs such as the data-usage statistics that have been exploited to infer a user's identity~\cite{DBLP:conf/ccs/ZhouDHNPWGN13} and to infer visited websites~\cite{DBLP:conf/wisec/SpreitzerGKM16}.

\attackSection{Empty Areas}
As can be observed, a few areas in this categorization system (cf. Figure~\ref{fig:big_picture}) are not (yet) covered or are not covered that densely. For instance, there is currently no active side-channel attack that can be executed remotely and that exploits logical properties (software features) to induce faults or to actively influence the targeted program (victim). 
However, by considering existing passive attacks, one could come up with more advanced attacks by introducing an active attacker. Such an active attacker might, for example, block/influence a shared resource in order to cause malfunctioning of the target. For instance, considering the passive attack exploiting the speaker status (on/off) to infer a user's driving routes~\cite{DBLP:conf/ccs/ZhouDHNPWGN13}, one could easily influence the victim application by playing inaudible sounds in the right moment in order to prevent the turn-by-turn voice guidance from accessing the speaker. Thereby, the active attacker prevents the target (victim) from accessing the shared resource, \ie, the speaker, and based on this induced behavior an active attacker might gain an advantage compared to a passive attacker. We expect advances in this (yet) uncovered area of active side-channel attacks that target software features. 

\attackSection{Tabular Summary of Surveyed Attacks}
Table~\ref{tab:summary_big_picture} provides a tabular summary for the categorization of the surveyed attacks. 
For some attacks we observe that active as well as passive modes of attack have already been considered, \eg, differential computation analysis and network traffic analysis attacks. 
Some attacks can also be conducted by exploiting physical properties as well as logical properties, \eg, the fingerprinting of devices and network traffic analysis attacks. 
\begingroup
\xpatchcmd\rxmark{\xmark}{ }{}{}
\setlength{\aboverulesep}{0pt}
\setlength{\belowrulesep}{0pt}
\begin{table*}
\centering
\caption{Summary of surveyed attacks.}
\label{tab:summary_big_picture}
\rowcolors{3}{white}{lightgray!40}
\begin{tabular}{lccccccc}
\toprule
\                                                 & \multicolumn{2}{c}{\textbf{Mode of attack}} & \multicolumn{2}{c}{\textbf{Exploited information}} & \multicolumn{3}{c}{\textbf{Location of attacker}} \\
\cmidrule(r{2pt}l{2pt}){2-3} \cmidrule(r{2pt}l{2pt}){4-5}  \cmidrule(r{2pt}l{2pt}){6-8} 
\multirow{-2}{*}{\rule{0pt}{2ex}\textbf{Attack}}  & \textbf{Active} & \textbf{Passive} & \textbf{Physical properties} & \textbf{Logical properties} & \textbf{Local} & \textbf{Vicinity} & \textbf{Remote} \\
\midrule
Power analysis                                    & \rxmark & \gcmark & \gcmark & \rxmark & \gcmark & \rxmark & \rxmark \\
EM analysis                                       & \rxmark & \gcmark & \gcmark & \rxmark & \gcmark & \gcmark & \rxmark \\
NAND mirroring                                    & \gcmark & \rxmark & \gcmark & \rxmark & \gcmark & \rxmark & \rxmark \\
Laser/optical                                     & \gcmark & \rxmark & \gcmark & \rxmark & \gcmark & \rxmark & \rxmark \\
Clock/power glitch                                & \gcmark & \rxmark & \gcmark & \rxmark & \gcmark & \rxmark & \rxmark \\
Temperature variation                             & \gcmark & \rxmark & \gcmark & \rxmark & \gcmark & \rxmark & \rxmark \\
EMFI                                              & \gcmark & \rxmark & \gcmark & \rxmark & \gcmark & \rxmark & \rxmark \\
Differential computation analysis                 & \gcmark & \gcmark & \rxmark & \gcmark & \gcmark & \rxmark & \rxmark \\
Reflections/hands                                 & \rxmark & \gcmark & \gcmark & \rxmark & \gcmark & \gcmark & \rxmark \\
Smudges                                           & \rxmark & \gcmark & \gcmark & \rxmark & \gcmark & \rxmark & \rxmark \\
Network traffic analysis                          & \gcmark & \gcmark & \gcmark & \gcmark & \rxmark & \gcmark & \rxmark \\
USB power analysis                                & \rxmark & \gcmark & \gcmark & \rxmark & \rxmark & \gcmark & \rxmark \\
Wi-Fi signal monitoring                            & \rxmark & \gcmark & \gcmark & \rxmark & \rxmark & \gcmark & \rxmark \\
Fingerprinting devices                            & \rxmark & \gcmark & \gcmark & \gcmark & \rxmark & \rxmark & \gcmark \\
Data-usage statistics                             & \rxmark & \gcmark & \rxmark & \gcmark & \rxmark & \rxmark & \gcmark \\
Page deduplication                                & \rxmark & \gcmark & \rxmark & \gcmark & \rxmark & \rxmark & \gcmark \\
procfs leaks                                      & \rxmark & \gcmark & \gcmark & \gcmark & \rxmark & \rxmark & \gcmark \\
Microarchitectural attacks                        & \gcmark & \gcmark & \gcmark & \rxmark & \rxmark & \rxmark & \gcmark \\
Location inference                                & \rxmark & \gcmark & \gcmark & \rxmark & \rxmark & \rxmark & \gcmark \\
Speech recognition                                & \rxmark & \gcmark & \gcmark & \rxmark & \rxmark & \rxmark & \gcmark \\
Soundcomber                                       & \rxmark & \gcmark & \gcmark & \rxmark & \rxmark & \rxmark & \gcmark \\
Sensor-based keyloggers                           & \rxmark & \gcmark & \gcmark & \rxmark & \rxmark & \rxmark & \gcmark \\
Rowhammer                                         & \gcmark & \rxmark & \gcmark & \rxmark & \rxmark & \rxmark & \gcmark \\
\bottomrule
\end{tabular} 
\end{table*}
\endgroup

\section{Discussion of Countermeasures}
 \label{sec:countermeasures}
In this section, we discuss existing countermeasures against the most prominent attacks. Overall we aim to shed light onto possible pitfalls of existing countermeasures and to stimulate future research to come up with more generic countermeasures against side-channel attacks. 

\subsection{Local Side-Channel Attacks}
\textbf{Protecting Cryptographic Implementations.}
Cryptographic implementations represent a prominent target of side-channel attacks as a successful attack allows to recover sensitive data and to break mechanisms building upon these primitives. Therefore, countermeasures to protect cryptographic implementations have already been proposed for the smart card world. These countermeasures can be applied to protect cryptographic implementations on smartphones as well. 
For example, masking of sensitive values such as the randomization of key-dependent values during cryptographic operations, or execution randomization are countermeasures for hardening the implementation against passive attacks such as power analysis or EM analysis \cite{DBLP:books/daglib/0017272}. Executing critical calculations twice allows to detect faults that are injected during an active side-channel attack \cite{DBLP:conf/fdtc/LomneRT12}. 

\attackSection{Protecting User Input}
Mitigation techniques to prevent attackers from inferring user input on touchscreens by means of smudge attacks or shoulder surfing attacks are not that thoroughly investigated yet. Nevertheless, proposed countermeasures include, for example, randomly starting the vibrator to prevent attacks that monitor the backside of the device~\cite{DBLP:conf/ndss/SunJCZZ016}, or to randomize the layout of the soft keyboard each time the user provides input to prevent smudge attacks~\cite{Aviv2012SideChannelsSmartphones} as well as attacks that monitor the hand movement~\cite{DBLP:conf/ccs/ShuklaKSP14}. Aviv~\cite{Aviv2012SideChannelsSmartphones} also proposed to align PIN digits in the middle of the screen and after each authentication the user needs to swipe down across all digits in order to hide smudges. Besides, Kwon and Na~\cite{DBLP:journals/compsec/KwonN14} introduced a new authentication mechanism denoted as \emph{TinyLock} that should prevent smudge attacks against pattern unlock mechanisms. Krombholz~\etal~\cite{DBLP:conf/soups/KrombholzHH16} proposed an authentication mechanism for devices with pressure-sensitive screens that should prevent smudge attacks and shoulder surfing attacks.  Raguram~\etal~\cite{DBLP:conf/ccs/RaguramWGMF11,DBLP:journals/tdsc/Raguram0XFGM13} suggested to decrease the screen brightness, to disable visual feedback (\eg, pop-up characters) on soft keyboards, and to use anti-reflective coating in eyeglasses to prevent attackers from exploiting reflections.

\subsection{Vicinity Side-Channel Attacks}
\textbf{Preventing Network Traffic Analysis.}
Countermeasures to prevent attackers from applying traffic analysis techniques on wiretapped network connections have been extensively considered in the context of website fingerprinting attacks. The main idea of these obfuscation techniques is to hide information that allows attackers to uniquely identify communication partners or transmitted content such as visited websites. 
Proposed countermeasures~\cite{DBLP:conf/ndss/WrightCM09,DBLP:conf/ndss/LuoZCLCP11,DBLP:conf/sp/DyerCRS12,DBLP:conf/wpes/CaiNJ14,DBLP:conf/wpes/NithyanandCJ14}, however, require the application, \eg, the browser application, as well as the remote server to cooperate. Furthermore, it has already been pointed out in~\cite{DBLP:conf/wisec/SpreitzerGKM16} that these countermeasures add overhead in terms of bandwidth and data consumption which might not be acceptable in case of mobile devices with limited data plans.

\subsection{Remote Side-Channel Attacks}
\textbf{Permissions.}
The most straight-forward approach always discussed as a viable means to prevent specific types of software-only side-channel attacks is to protect the exploited information or resource by means of dedicated permissions. However, there is a study~\cite{DBLP:conf/soups/FeltHEHCW12} that showed that permission-based approaches are not quite convincing. Some users do not understand the exact meaning of specific permissions, and others do not care about requested permissions. Acar~\etal~\cite{DBLP:conf/sp/Acar0BFM016} even attested that the Android permission system ``has failed in practice''. Despite these problems it seems to be nearly impossible to add dedicated permissions for every exploited information. 

\attackSection{Keyboard Layout Randomization}
In order to prevent sensor-based keylogging attacks that exploit the correlation between user input and the device movements observed via sensor readings, the keyboard layout of soft keyboards could be randomized~\cite{DBLP:conf/wmcsa/OwusuHDPZ12}. 
For instance, the Android-based CyanogenMod OS allows to enable such a feature for PIN inputs optionally. However, it remains an open question how this would affect usability in case of QWERTY keyboards and, intuitively, it might make keyboard input nearly impossible. 

\attackSection{Limiting Access or Sampling Frequency}
It has also been suggested to disable access to sensor readings during sensitive input or to reduce the sampling frequency of sensors. This, however, would hinder applications that heavily rely on sensor readings such as pedometers. 

Side-channel attacks like \emph{Soundcomber} might be prevented by \emph{AuDroid}~\cite{DBLP:conf/acsac/PetraccaSJA15}, which is an extension to the SELinux reference monitor that has been integrated into Android to control access to system audio resources. As pointed out by the authors, there is no security mechanism in place for the host OS to control access to mobile device speakers, thus allowing untrusted apps to exploit this communication channel. \emph{AuDroid} enforces security policies that prevent data in system apps and services from being leaked to (or used by) untrusted parties. 

\attackSection{Noise Injection}
Randomly starting the phone vibrator has been suggested by Owusu~\etal~\cite{DBLP:conf/wmcsa/OwusuHDPZ12} to prevent sensor-based keyloggers that exploit the accelerometer sensor. However, Shrestha~\etal~\cite{DBLP:conf/wisec/ShresthaMS16} showed that random vibrations do not provide protection. As an alternative, Shrestha~\etal~proposed a tool named \emph{Slogger} that introduces noise into sensor readings as soon as the soft keyboard is running. In order to do so, Slogger relies on a tool that needs to be started via the ADB shell (in order to be executed with ADB capabilities). Slogger injects events into the files corresponding to the accelerometer and the gyroscope located in \verb|/dev/input/|, which is why ADB privileges are required for this defense mechanism. The authors even evaluated the effectiveness of Slogger against two sensor-based keyloggers and found that the accuracy of sensor-based keyloggers can be reduced significantly. Das~\etal~\cite{DBLP:conf/ndss/DasBC16} also suggested to add noise to sensor readings in order to prevent device fingerprinting via hardware imperfections of sensors. A more general approach that targets the injection of noise into the information provided via the procfs has been proposed by Xiao~\etal~\cite{DBLP:conf/ccs/XiaoRZ15}. 

\attackSection{Preventing Microarchitectural Attacks}
The inherent nature of modern computer architectures enables sophisticated attacks due to shared resources and especially due to dedicated performance optimization techniques. A famous and popular example is the memory hierarchy that introduces significant performance gains but also enables microarchitectural attacks such as cache attacks. Although specific cryptographic implementations can be protected against such attacks, \eg, bit-sliced implementations~\cite{DBLP:conf/ctrsa/Konighofer08,DBLP:conf/cans/RebeiroSD06} or dedicated hardware instructions can be used to protect AES implementations, generic countermeasures against cache attacks represent a non-trivial challenge. However, we consider it of utmost importance to spur further research in the context of countermeasures, especially since cache attacks do not only pose a risk for cryptographic algorithms, but also for other sensitive information such as keystrokes~\cite{DBLP:conf/uss/GrussSM15,DBLP:conf/uss/LippGSMM16}. 

\attackSection{App Guardian}
Most of the above presented countermeasures aim to prevent very specific attacks only, but cannot be applied to prevent attacks within a specific category of our classification system, \eg, software-only attacks located in the upper right of our new classification system (cf. Figure~\ref{fig:big_picture}). At least some of these attacks, however, have been addressed by App Guardian~\cite{DBLP:conf/sp/ZhangY0ZW15}, which represents a more general approach to defend against software-only attacks. App Guardian is a third-party application that runs in user mode and employs side-channel information to detect RIG attacks (including software-only side-channel attacks). The basic idea of App Guardian is to stop the malicious application while the principal (the app to be protected) is being executed and to resume the (potentially malicious) application later on. Although App Guardian still faces challenges, it is a novel idea to cope with such side-channel attacks in general. More specifically, it tries to cope with all passive attacks that require the attacker to execute software on the targeted device. 

App Guardian seems to be a promising research project to cope with side-channel attacks on smartphones at a larger scale. However, an unsolved issue of App Guardian is the problem that it still struggles with the proper identification of applications to be protected.
Furthermore, App Guardian relies on side-channel information---to detect ongoing side-channel attacks---that has been removed in Android 7. 
Hence, App Guardian needs to be updated in order to also work on recent Android versions and its effectiveness should be further evaluated against existing side-channel attacks. Furthermore, it might be interesting to extend its current framework to cope with side-channel attacks conducted from within the browser, \ie, to mitigate side-channel attacks via JavaScript.

\subsection{Summary}
Although local attacks target only a few devices or users, we also observe that we require a much broader range of countermeasures because also the attack methodologies of local attacks are much broader. 
For instance, we have to deal with attackers that measure the power consumption of the device in order to break cryptographic implementations, we have to deal with fault attacks such as clock/power glitching and temperature variations, and at the same time we have to deal with attackers that exploit smudges left on the touchscreen. 

In contrast, the commonality of all remote attacks is that they require software execution on the targeted device. 
Although this means that remote attacks target devices and users at a much broader scale, more generic countermeasures such as App Guardian seem to be the most promising approach to cope with these attacks in the future.

\section{Issues, Challenges, and Future Research}
\label{sec:future_research}
In this section we discuss open issues and challenges that need to be addressed in future research. Hence, this section is not meant to provide solutions to existing problems. Instead, with the presented classification system for modern side-channel attacks we aim to shed light onto this vivid research area and, thereby, to point out high-level research directions. Overall, the ultimate goal is to spur further research in the context of side-channel attacks and countermeasures and, as a result, to pave the way for a more secure computing platform for smart and mobile devices. 

\attackSection{Countermeasures}
Side-channel attacks are published at an unprecedented pace and appropriate defense mechanisms are often either not (yet) available or cannot be deployed easily. Especially the five \emph{key enablers} identified in this paper enable devastating side-channel attacks that can be conducted remotely and, thus, target an immense number of devices and users at the same time. Although countermeasures are being researched, we observe a cat and mouse game between attackers and system engineers trying to make systems secure from a side-channel perspective. Besides, even if effective countermeasures were readily available, the mobile ecosystem of Android would impede a large-scale deployment of many of these defense mechanisms. The main problem is that even in case Google would apply defense mechanisms and patch these information leaks, multiple device manufacturers as well as carriers also need to apply these patches to deploy countermeasures in practice. Hence, chances are that such countermeasures will never be deployed, especially not in case of outdated operating systems. We hope to stimulate research to come up with viable countermeasures in order to prevent such attacks at a larger scale, \ie, by considering larger areas within the new categorization system, while also considering the challenges faced by the mobile ecosystem. For instance, App Guardian~\cite{DBLP:conf/sp/ZhangY0ZW15} follows the right direction by trying to cope with attacks at a larger scale, while at the same time it can be deployed as a third-party application. 

\attackSection{Reproducibility and Responsible Disclosure}
In order to foster research in the context of countermeasures, it would be helpful to publish the corresponding frameworks used to conduct side-channel attacks. While this might also address the long-standing problem of reproducibility of experiments in computer science in general, this would enable a more efficient evaluation of developed countermeasures. At the same time, however, responsible disclosure must be upheld, which sometimes turns out to be a difficult balancing act. On the one hand, researchers want to publish their findings as soon as possible and on the other hand, putting countermeasures to practice might take some time. 

\attackSection{Different Mobile Operating Systems and Cross-Platform Development}
Research should not only focus on one particular OS exclusively, \ie, especially Android seems to attract the most attention. Instead, the applicability of side-channel attacks should be investigated on multiple platforms, as many (or most) of the existing attacks work on other platforms as well. This is due to the fact that different platforms and devices from different vendors aim to provide the same features such as sensors and software interfaces, and rely on similar security concepts like permission systems and application sandboxing. 

In addition, the increasing trend to develop applications for multiple platforms (cross-platform development) also increases the possibility to target multiple platforms at the same time. For example, the increasing popularity of HTML5 apps and the increasing availability of web APIs to access native resources from JavaScript significantly increases the scale of side-channel attacks as specific attacks possibly target multiple platforms at the same time. 

\attackSection{Wearables}
Although we put a strong focus on smartphones in this paper, we stress that wearables in general must be considered in future research. For example, smartwatches have already been employed to attack user input on POS terminals and hardware QWERTY keyboards~\cite{DBLP:conf/mobicom/WangLC15,DBLP:conf/ccs/LiuZDLZ15,DBLP:conf/ccs/MaitiAJH16,DBLP:conf/ccs/WangGWCL16}. Besides, it has also been demonstrated that smartwatches can be used to infer input on smartphones~\cite{DBLP:conf/wifs/SarkisyanDN15,DBLP:conf/iswc/MaitiJHB15} as well as text written on whiteboards~\cite{DBLP:conf/percom/ArduserBBW16}. With the ever increasing number of smart devices connected to our everyday lives, the threat of side-channel attacks increases. We are likely to see higher accuracies when these attacks are performed across multiple devices, \eg, when combining data from smartwatches and smartphones. Furthermore, Farshteindiker~\etal~\cite{DBLP:conf/woot/FarshteindikerH16} also demonstrated how hardware implants (bugs)---possibly used by intelligence agencies---can be used to exfiltrate data by communicating with a smartphone. The communication channel is based on inaudible sounds emitted from the implant which are captured by the gyroscope of the smartphone. This interconnection clearly demonstrates the potential of attacks when multiple wearable devices are combined.

\attackSection{Internet of Things}
Another area of research which is rapidly growing is the Internet of Things (IoT). As all devices in the IoT network are inter-connected and accessible via the Internet, we foresee that attackers will exploit side-channel leaks to target different kinds of IoT appliances. In fact such an attack has already been carried out by Zhang~\etal~\cite{DBLP:conf/sp/ZhangY0ZW15}. They investigated an Android-based Wi-Fi camera and observed that a particular side-channel leak on Android can be exploited to infer whether or not the user is at home. This example demonstrates that side-channel leaks do not only pose a threat to a user's privacy and security from a system security point of view, but also pose a threat to smart home appliances and security systems, such as smart thermostats, cameras, and alarm systems. Although this sounds utopian at first, the above example clearly demonstrates that side-channel leaks (on smartphones) also pose a threat to these IoT appliances and puts even users' physical possessions at risk.

\attackSection{Combination of Multiple Information Leaks}
In order to improve the accuracy of existing attacks or to come up with more sophisticated attack scenarios, multiple side-channel leaks can also be combined. For instance, the combination of cache attacks and sensor-based keyloggers as mentioned in~\cite{DBLP:conf/uss/LippGSMM16} could be used to improve keylogging attacks. First, cache attacks can be used to determine the exact time when a key is entered and, second, sensor-based keyloggers can be used to infer the actual key. 
Furthermore, website fingerprinting attacks could be combined with sensor-based keyloggers as mentioned in~\cite{DBLP:conf/wisec/SpreitzerGKM16}, which would allow to steal login credentials for specific websites. 

In addition, side-channel attacks can also be used to improve attacks that exploit software vulnerabilities. For example, although Screenmilker~\cite{DBLP:conf/ndss/LinLZW14} does not represent a side-channel attack---because a software vulnerability is exploited---it relies on side-channel information in order to exploit this vulnerability in the right moment. Lin~\etal~\cite{DBLP:conf/ndss/LinLZW14} suggested to rely on CPU utilization, memory consumption, and network activities in order to determine whether the targeted app is executed and, thus, were able to take screenshots in the right moment.  

\attackSection{Code Analysis Tools}
The appification of mobile devices enables an easy installation of apps from the app markets. However, these apps can be implemented by anyone who has a developer account and, thus, the code needs to be checked and verified appropriately, \ie, for presence of malicious behavior and side channels. While the app vetting processes of app stores, \eg, Google Play, already check for presence of malicious behavior, dedicated technologies, such as static and dynamic code analysis, should also be employed in order to prevent apps prone to side-channel attacks and apps exploiting side-channel information leaks from being distributed via app markets. This, however, does not seem to be a trivial task since most side-channel attacks exploit information or resources that can be accessed without any specific privileges or permissions. 

Static and dynamic code analysis tools could also help to fix implementation flaws that lead to side-channel attacks. 
Some implementation flaws exist for many years without being noticed as has been demonstrated in \cite{DBLP:conf/ccs/GarciaBY16} for the OpenSSL implementation of the digital signature algorithm. Fostering the development and application of tools to find and detect such flaws during the software development process could help to prevent vulnerable code from being deployed.

A possible starting point for the investigation and extension of code analysis tools that might allow to scan applications for possible side-channel attacks would be one of the survey papers discussed in Section~\ref{sec:related_surveys}.

\section{Conclusion}
 \label{sec:conclusion}
Considering the immense threat arising from side-channel attacks on mobile devices, a thorough understanding of information leaks and possible exploitation techniques is necessary. Based on this open issue, we surveyed existing side-channel attacks and identified commonalities between these attacks in order to systematically categorize all existing attacks. With the presented classification system we aim to provide a thorough understanding of information leaks and hope to spur further research in the context of side-channel attacks as well as countermeasures and, thereby, to pave the way for secure computing platforms.

\section*{Acknowledgment}
The research leading to these results has received funding from the European Union's Horizon 2020 research 
\begin{wrapfigure}{l}{4.5cm}
    \begin{tabular}{@{}c@{}}
      \includegraphics[width=4.5cm]{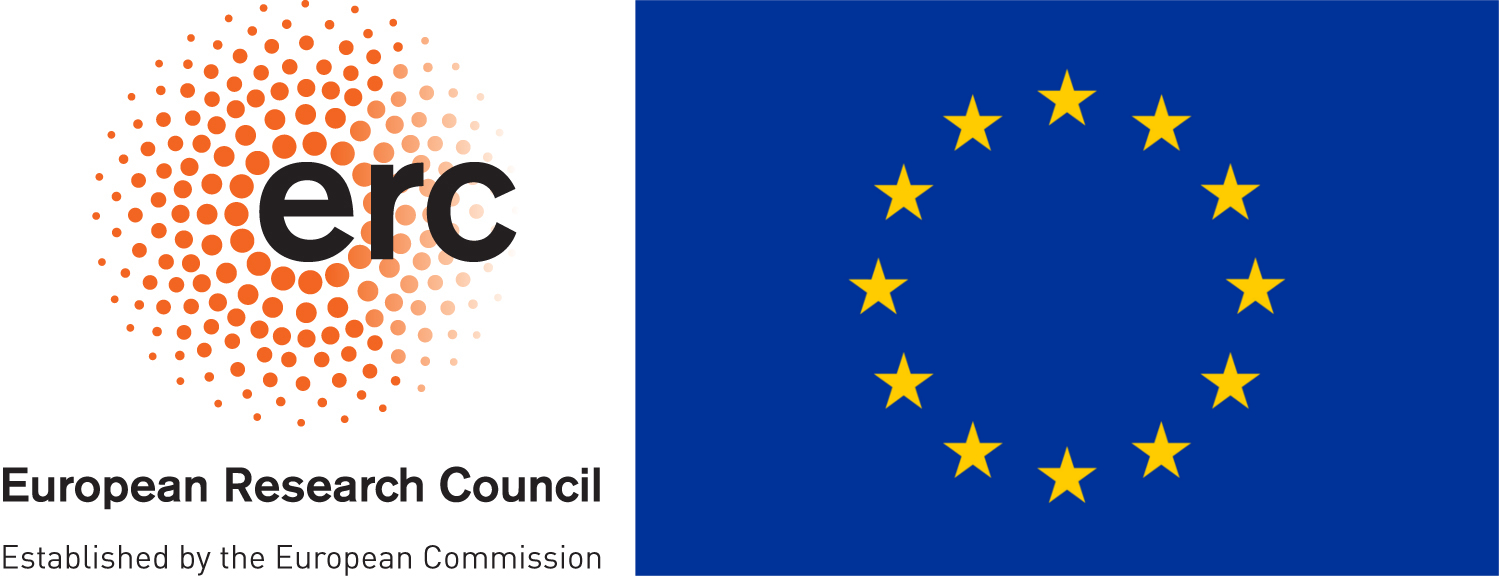}
    \end{tabular}
\end{wrapfigure}
and innovation programme 
under grant agreement 
No 644052 (HECTOR),
and the European Research Council (ERC) under the European Union's Horizon 2020 research and innovation programme (grant agreement No 681402).
Veelasha Moonsamy has been supported by the Technology Foundation STW (project 13499 - TYPHOON \& ASPASIA) from the Dutch government.
Further, we would like to thank Florian Mendel for helpful discussions about active side-channel attacks as well as Cristofaro Mune and Nikita Abdullin for pointing out  a missing attack category.


\ifCLASSOPTIONcaptionsoff
  \newpage
\fi



\bibliographystyle{IEEEtran}
\bibliography{bibliography}
%



%

\begin{IEEEbiography}[{\includegraphics[width=1in,height=1.25in,clip,keepaspectratio]{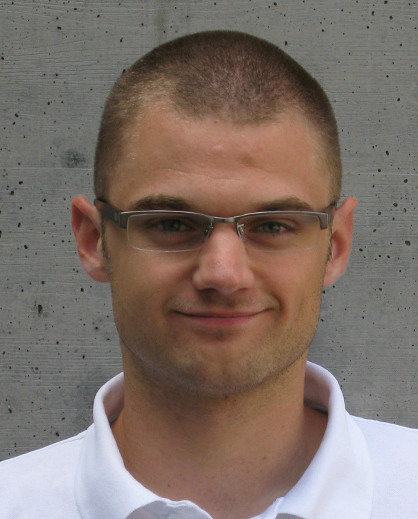}}]{Raphael Spreitzer}
is a researcher at Graz University of Technology. His main research interests are information security with a special focus on side-channel attacks on mobile devices, \eg, cache attacks and sensor-based attacks, and practical applications of privacy-enhancing technologies. He obtained a PhD degree in computer science with distinction from Graz University of Technology in 2017. Before, he finished the master's programme Software Engineering and Management with distinction at Graz University of Technology.
\end{IEEEbiography}

\begin{IEEEbiography}[{\includegraphics[width=1in,height=1.25in,clip,keepaspectratio]{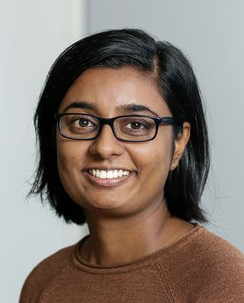}}]{Veelasha Moonsamy}
is a postdoctoral researcher in the Digital Security group at Radboud University in The Netherlands. She obtained her PhD from Deakin University in Melbourne (Australia), under the supervision of Prof. Lynn Batten. Her research interests revolves around security and privacy on mobile devices, in particular side- and covert-channel attacks, malware detection and mitigation of information leaks at application and hardware level.
\end{IEEEbiography}


\begin{IEEEbiography}[{\includegraphics[width=1in,height=1.25in,clip,keepaspectratio]{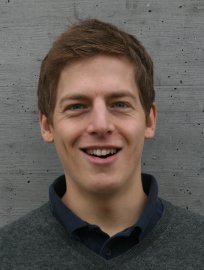}}]{Thomas Korak}
received his MSc (Dipl.-Ing.) in Computer Engineering at Graz University of Technology in 2011 and his PhD in 2015. 
Until 2017 he was a postdoctoral researcher at the Institute for Applied Information Processing and Communications (IAIK), Graz University of Technology. His main research topics are side-channel attacks and fault attacks targeting embedded devices. Next to that he is also interested in countermeasures for hardening devices against this kind of attacks. 
\end{IEEEbiography}

\begin{IEEEbiography}[{\includegraphics[width=1in,height=1.25in,clip,keepaspectratio]{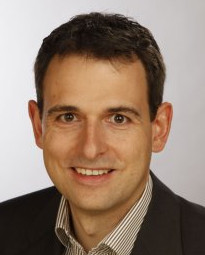}}]{Stefan Mangard}
is full professor at Graz University of Technology since November
2013. Before moving to Graz, he was working as leading security architect at Infineon
Technologies in Munich. In this role he was responsible for defining the security concepts for
all the smart card platforms of Infineon, one of the largest manufacturers of smart card ICs
worldwide.
He is chair of the steering committee of CHES, which is the foremost conference on
cryptographic hardware, and associate editor of the Journal on Cryptographic Engineering.
He regularly serves on program committees of conferences in the field (CHES, CARDIS,
COSADE...). In 2015, Stefan has received the highly prestigious ERC consolidator grant by the European Research Council for his research proposal ``SOPHIA -- Securing Software against Physical Attacks''.
\end{IEEEbiography}


\vfill


\end{document}